\documentclass[%
 reprint,
superscriptaddress,
 amsmath,amssymb,
 aps,
pra,
]{revtex4-2}
\usepackage{physics} 
\usepackage{graphicx}% Include figure files
\usepackage{dcolumn}% Align table columns on decimal point
\usepackage{bm}% bold math
\usepackage{xcolor}

\usepackage{lipsum}
\usepackage[english]{babel}

\usepackage[colorlinks=true,
            linkcolor=blue,
            citecolor=blue,
            urlcolor=blue]{hyperref}

\begin{document}

\preprint{APS/123-QED}

\title{Noise Correlations as a Resource in Pauli-Twirled Circuits}% Force line breaks with \\

\author{Antoine Brillant}
\affiliation{Pritzker School of Molecular Engineering, University of Chicago, Chicago, IL, USA}

\author{Rohan N Rajmohan}
\affiliation{Department of Physics and Astronomy, Northwestern University, Evanston, Illinois 60208, USA}

\author{Peter Groszkowski}
\affiliation{National Center for Computational Sciences, Oak Ridge National Laboratory, Oak Ridge, TN, USA}

\author{Alireza Seif}
\affiliation{IBM Quantum, IBM T.J. Watson Research Center, Yorktown Heights, NY, USA}

\author{Jens Koch}
\affiliation{Department of Physics and Astronomy, Northwestern University, Evanston, Illinois 60208, USA}

\author{Aashish Clerk}
\affiliation{Pritzker School of Molecular Engineering, University of Chicago, Chicago, IL, USA}

\date{\today}% It is always \today, today,

\begin{abstract}
Randomized compiling (RC) is an established tool to tailor arbitrary quantum noise channels into Pauli errors. The effect of both spatial and temporal noise correlations in randomly compiled circuits, however, is not fully understood.  Here, we show that for a broad class of correlated Gaussian noise, RC reduces both the strength and temporal range of correlations. For Clifford circuits, we derive a simple analytical expression for the circuit fidelity of randomly compiled circuits. Surprisingly, we show that this fidelity is always increased by the presence of correlations, suggesting that correlations are a resource in randomly compiled circuits. To leading order in system-bath coupling, we also show that RC suppresses the quantum component of bath correlations, implying that one can safely treat weak noise as being classical. Finally, through extensive numerical simulations, we show that our results remain valid for many relevant non-Clifford circuits. These results clarify how RC mitigates memory effects and enhances circuit robustness.
\end{abstract}
% Its effect on temporally correlated errors, however, is not fully understood.
\maketitle

%\tableofcontents

\section{Introduction}
\label{sec:intorduction}

Accurately understanding the effect of noise on quantum circuits is becoming increasingly important as devices scale in size and complexity. Generic error processes are characterized by a number of parameters that grows exponentially with system size, making them notoriously hard to characterize, simulate and mitigate \cite{cai_quantum_2023}. In contrast, Pauli noise models can be characterized by a relatively small number of parameters when additional structure -- such as locality and sparsity -- are present \cite{vandenbergProbabilisticErrorCancellation2023}.

These noise models can be efficiently simulated for Clifford circuits using the stabilizer formalism. They also unify common error metrics: for Pauli channels, the average gate fidelity \cite{nielsenSimpleFormulaAverage2002} and the diamond norm \cite{aharonov_quantum_1998} are equivalent, making standard learning protocols such as randomized benchmarking \cite{ryanRandomizedBenchmarkingSingle2009, magesanCharacterizingQuantumGates2012, epsteinInvestigatingLimitsRandomized2014} or cycle benchmarking \cite{erhardCharacterizingLargescaleQuantum2019} sufficient to fully characterize their behavior.

\begin{figure}[h!]
    \centering
    \includegraphics[width=1\linewidth]{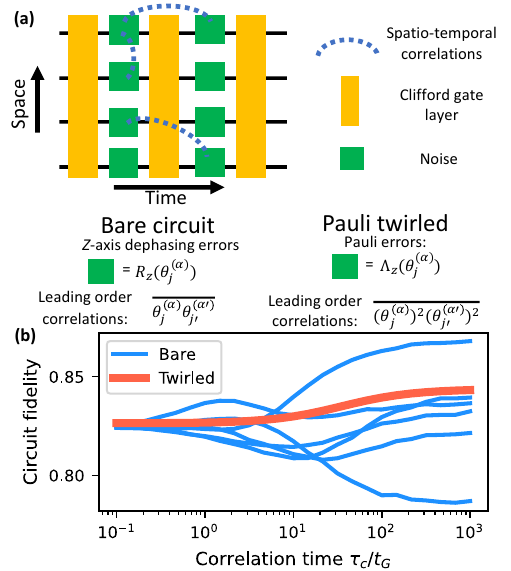}
    \caption{
    %Schematic representation of the main results of this paper. 
    (a)  A quantum circuit composed of layers of Clifford gates (yellow rectangle) subject to spatio-temporally correlated noise (green boxes with dotted lines). We consider two noise models, correlated dephasing along the $Z$ axis and correlated Pauli channels [c.f. Eq.~\eqref{overrotations} and Eq.~\eqref{pauli_layer}]. (b) Circuit fidelity of bare and twirled two-qubit Clifford circuits as a function the correlation time of the noise. The circuits are composed of $50$ layers of random two-qubit Clifford gates. Noise temporal correlations are decaying exponentially with correlation time $\tau_c$ and strength $\sigma = 0.05$ and is spatially uncorrelated [c.f. Eq.~\eqref{corr_func}]. Each blue curve corresponds to a different bare Clifford circuit fidelity which are equivalent after twirling the noise (red curve). The twirled circuit fidelity increases with the correlation time of the noise. 
    }
    
    % Twirling transforms correlated stochastic coherent errors into correlated stochastic Pauli channels, reducing the effective strength of the noise correlations. The dependence of the circuit on the average circuit fidelity is then greatly reduced.
    \label{fig:cartoon_twirl_vs_untwirl}
\end{figure}

Although Pauli noise models generally do not naturally arise on quantum devices, they remain relevant. For example, under mild assumptions, protocols such as randomized compiling (RC) can tailor general error channels into Pauli channels by randomizing over equivalent circuit decompositions \cite{wallmanNoiseTailoringScalable2016, winickConceptsConditionsError2022, hashim_randomized_2021}. Averaging over these random compilations yields an average noise model that effectively corresponds to Pauli noise. In addition to simplifying characterization and simulation, tailoring arbitrary error channels into Pauli channels also suppresses coherent errors, which can be severely detrimental to circuit fidelity \cite{hashim_randomized_2021, wallmanEstimatingCoherenceNoise2015, sanders_bounding_2015, winickConceptsConditionsError2022}.

While RC is well understood in the absence of noise correlations, important questions remain about its efficacy in the more realistic case of correlated noise.  In particular, RC has been shown to suppress temporal correlations through a mechanism akin to randomized dynamical decoupling \cite{winickConceptsConditionsError2022}. 
However, the cause and consequences of the residual correlations that survive RC are largely unexplored.

% How are they connected to the underlying microscopic properties of relevant noise processes? How do they affect performance? In the presence of noise correlations,  can errors interfere constructively and enhance fidelity metrics, or destructively and significantly harm them? Perhaps more importantly, should we worry about spatio-temporal correlations when applying RC to realistic devices?

In this paper, we build upon a recently introduced analytical framework \cite{brillant_randomized_2025} to further the understanding of RC and correlated noise. We study the circuit fidelity of Pauli-twirled Clifford circuits subject to classical, Gaussian non-Markovian noise (also possibly spatially-correlated) and derive tight bounds using a generalized cumulant expansion. Surprisingly, after twirling the noise, the lower bound on the circuit fidelity of any given Clifford circuit with fixed single gate error rate, corresponds to the limit of Markovian noise with no spatial correlations. On the other hand, the upper bound is given by the opposite limit of noise  maximally correlated in space and time. This implies that noise correlations can be seen as a resource in RC circuits (see Fig.~\ref{fig:cartoon_twirl_vs_untwirl}). We note that correlation-enhanced performance has been observed in related settings for variational algorithms \cite{kattemolleAbilityErrorCorrelations2023}. Our work extends this picture by establishing rigorously and non-perturbatively that correlations are strictly beneficial in Pauli-twirled circuits for a broad class of Gaussian noise models.

While RC is often beneficial, suppressing coherent error buildup can also remove potentially beneficial destructive interference. Nevertheless, even when it is unclear whether RC will improve the fidelity of a given circuit, it remains useful. We observe that RC reduces circuit-to-circuit variations in fidelity, making the impact of the noise far more predictable. In contrast, predicting the effect of the original noise on a circuit is generally much more difficult.

Our analytical results focus on Clifford circuits subject to classical noise acting between idealized gate layers, i.e., neglecting any interplay between noise and gate implementation. We now comment on extensions beyond this setting. While correlated Pauli noise can always be described by an effective classical noise model (see e.g. \cite{liuNonMarkovianNoiseSuppression2024a}), this description may still reflect properties of the underlying quantum environment. Here, we show that RC strongly suppresses the leading quantum component of noise correlations, further justifying a classical treatment. Our framework also extends to finite-duration gates, where noise acts during the gate and couples to its implementation. In this case, we obtain similar bounds on the circuit fidelity. Extensions to non-Clifford circuits and adaptive protocols fall outside our analytical framework. We therefore test our predictions numerically on non-Clifford algorithms and error-correction circuits, including an 18-qubit quantum Fourier transform, a square-root amplitude amplification circuit, and a three-qubit repetition code with mid-circuit measurements and feedback. We observe behavior consistent with our analytical predictions: the fidelity of twirled circuits increases with the amount of temporal correlations, suggesting that our conclusions remain valid in experimentally relevant settings.

The rest of this paper is organized as follows, in Sec.~\ref{sec:main results}, we derive our bound in the case of Clifford circuits subject to classical dephasing noise and show how RC reduces the effect of temporal correlations. In Sec.~\ref{sec:quantum-finite-gates}, we show that our result partially extends to finite-duration gate implementations and we analyze the effect of RC on quantum correlations. In Sec.~\ref{sec:applications}, we analyze near-term, non-Clifford circuits through numerical simulations, and show that in many relevant situations, our results remain valid for non-Clifford circuits. Finally, we conclude in Sec.~\ref{sec:discussion}.

\section{Main results}
\label{sec:main results}

In this section, we analyze the effect of classical noise correlations on Clifford circuits that are randomly compiled. To do so, we focus on the circuit fidelity:
\begin{equation}
    F_{\text{tot}} = \int d\psi ~ \mathrm{tr} \big[\mathcal U_{\text{ideal}}(\dyad{\psi}) \mathcal U_{\text{noisy}}(\dyad{\psi})\big],
\end{equation}
where $\mathcal U_{\text{ideal}}$ and $\mathcal U_{\text{noisy}}$ are the superoperators implementing the target circuit and its noisy versions (c.f. Eqs.~\eqref{U_ideal}-\eqref{noisy_prop}). We derive a simple analytical formula for the fidelity of such circuits and use it to obtain tight bounds, showing that increasing the amount of spatio-temporal noise correlation generically improves the fidelity in twirled circuits. % Finally, we show that while the effect of correlations is positive in randomly compiled circuits, the magnitude of these correlations is reduced by RC, implying that circuits designed to dynamically decouple the noise (see, e.g., \cite{violaDynamicalDecouplingOpen1999}) from the system could suffer from RC. However, we stress that most practical circuits do not have that property, making RC generically useful (see e.g Refs.~\cite{wallmanNoiseTailoringScalable2016, winickConceptsConditionsError2022, hashim_randomized_2021} and Sec.~\ref{sec:applications}).

\subsection{General setting: Pauli-twirled Clifford circuits with correlated noise}
\label{subsec:model}

We consider an $n$-qubit quantum circuit composed of $l$ layers of Clifford gates.  
In the absence of noise, the entire circuit can be represented by the superoperator
\begin{equation}
    \mathcal{U}_{\text{ideal}} = \mathcal{G}_l \dots  \mathcal{G}_1, \label{U_ideal}
\end{equation}
where each $\mathcal{G}_j[\cdot] = \hat{g}_j [\cdot] \hat{g}_j^{\dagger}$ denotes the unitary superoperator corresponding to the application of the $j$th Clifford layer $\hat{g}_j$.  

Noise will of course disrupt this ideal unitary evolution. We consider the case of classical noise coupling to the qubits’ $Z$ axes, described by the Pauli operators $\hat{\sigma}_z^{(\alpha)}$ where $\alpha$ indexes the qubits. The assumption of $Z$-axis noise is made for clarity, but can be relaxed to include noise that couples to arbitrary system operators, as discussed below. Following each unitary Clifford layer, qubit $\alpha$ experiences dephasing along the Z-axis
\begin{equation}
    \hat{R}_{Z^{(\alpha)}}(\theta_j^{(\alpha)}) 
    = \exp\!\left[-i\hat{\sigma}_z^{(\alpha)}\theta_j^{(\alpha)}\right], \label{overrotations}
\end{equation}
where $\theta_j^{(\alpha)}$ are random error angles that depend on both the layer index $j$ and the qubit index $\alpha$. Physically, this model corresponds to treating each gate layer as instantaneous, followed by a noisy idle period. This simplification clarifies the presentation of our main results. As shown below in Section \ref{sec:quantum-finite-gates}, incorporating finite-duration gates does not modify our main conclusions.
Our focus is understanding the impact of temporal and spatial correlations in the noise. We thus describe the random variables $\theta_j^{(\alpha)}$ to be correlated in both their spatial and temporal indices $j$ and $\alpha$. At this stage, details of their distribution are unimportant and we will specify it when needed.

We group all the error angles acting in a given layer into the vectors $\vec \theta_j = (\theta_j^{(1)}, \dots, \theta_j^{(n)})$ and define the superoperators that apply a full layer of errors:
\begin{align}
    \mathcal{R}_Z(\vec{\theta}_j) 
    &= \prod_{\alpha=1}^n \mathcal{R}_Z^{(\alpha)}(\theta_j^{(\alpha)}),\\
    \qquad 
    \mathcal{R}_Z^{(\alpha)}(\theta_j^{(\alpha)})[\cdot]
    &= \hat{R}_Z^{(\alpha)}(\theta_j^{(\alpha)})[\cdot]
      \hat{R}_Z^{(\alpha)\dagger}(\theta_j^{(\alpha)}).
\end{align}
For a given realization of the random error angles, the noisy circuit is represented by the superoperator
\begin{equation}
    \mathcal{U}_{\text{noisy}}(\vec{\theta}_1, \dots, \vec{\theta}_l)
    = \mathcal{R}_Z(\vec{\theta}_l)\mathcal{G}_l 
   \dots  
    \mathcal{R}_Z(\vec{\theta}_1)\mathcal{G}_1, \label{noisy_prop}
\end{equation}
which we will eventually average over the noise. Note that for any particular noise realization, the errors are coherent and can add up constructively as the circuit depth increases.  
In the worst case, this can lead to error rates that grow quadratically with the number of layers, rather than linearly.

A powerful method for preventing this buildup of coherent errors is RC (which we will also refer to as Pauli-twirling) \cite{wallmanNoiseTailoringScalable2016}.  
This protocol takes advantage of the fact that many circuits are logically equivalent: they implement the same unitary but differ in their intermediate gate sequences.  
By sampling from a distribution of such equivalent circuits and averaging over their outcomes, one can effectively tailor the noise into stochastic Pauli errors.  
In the case of Clifford circuits, this randomization can be achieved in a particularly simple way. First, insert a layer of random $n$-qubit Pauli gates $\hat{P}_j$ before each Clifford layer. Second, apply the Pauli gates $\hat{P}_j^c$ that undo the effect of random Pauli after the layer. In their superoperator form these inverses are defined by:
\begin{equation}
    \mathcal{P}_j^c = \mathcal{G}_j \mathcal{P}_j \mathcal{G}_j^{-1},
\end{equation}
where we defined the unitary superoperators $\mathcal P_j = \hat P_j [\cdot] \hat P_j$ implementing the Pauli gates. Because $\mathcal{P}_j^c$ is chosen to compensate for $\mathcal{P}_j$ under the Clifford action, the overall logical circuit remains unchanged.
In other words, the ideal computation is unaffected, while the physical noise experienced by the qubits is randomized. This randomization is due to the fact that the errors do not generally commute with the inserted Pauli operators. The effective noisy layer then becomes:
\begin{equation}
    \mathcal{P}_j^c \mathcal{R}_Z(\vec{\theta}_j) \mathcal{G}_j \mathcal{P}_j
    = \mathcal{P}_j^c \mathcal{R}_Z(\vec{\theta}_j) \mathcal{P}_j^c \mathcal{G}_j. \label{twirled_channel}
\end{equation}
Here we assume that the Pauli gates themselves are noiseless, since they can typically be absorbed into existing single qubit gates at no additional cost.

At this point, we notice that our circuit has been randomized in two different ways: both by the intrinsic circuit noise, as well as by our deliberate insertion of random Pauli gates.  We are interested in understanding the circuit fidelity averaged over both of these randomizations. We will proceed by first averaging over the random Pauli gates, and only then doing a noise average. As we will see, this choice will be useful for two reasons. First, Pauli-twirling transforms the stochastic coherent error channels into stochastic Pauli channels (with random error probabilities that depend on the error angles), which are easier to manipulate than coherent error channels. Second, Pauli-twirling significantly changes how long range temporal correlations act on the system. This simplified structure will be enough to treat the effect of noise correlations non-perturbatively in the correlation time of the noise. This is remarkable as for many problems involving non-Markovian noise (including classical colored noise), analytic solutions beyond short-time perturbation theory are rarely available, and one typically resorts to methods such as generalized cumulant expansions \cite{groszkowskiSimpleMasterEquations2023, lidar_lecture_2020, foxCritiqueGeneralizedCumulant1976}. In contrast, our setting permits a non-perturbative treatment of temporal correlations.

\subsection{Stochastic averaging}

\subsubsection{Averaging over random Paulis}

The first step in our analysis is thus to average over the random Pauli gates. Since the inserted Pauli gates are uncorrelated between different layers, we can average each layer independently. Denoting the average over Paulis by $\langle \cdot \rangle_{\mathcal P} \equiv \frac{1}{4^n} \sum_{\mathcal P} [\cdot]$, one finds
\begin{align}
    \langle \mathcal{P}\, \mathcal{R}_Z(\vec{\theta}_j)\, \mathcal{P} \rangle_{\mathcal P}
    &= \prod_{\alpha=1}^{n}
        \Big[ \cos^2(\theta_j^{(\alpha)})\, \mathcal{I}
        + \sin^2(\theta_j^{(\alpha)})\, \mathcal{Z}^{(\alpha)} \Big] \label{eq:twirled_layer}\\
    &\equiv \prod_{\alpha=1}^{n} \Lambda_{Z^{(\alpha)}}(\theta_j^{(\alpha)}), \label{pauli_layer}
\end{align}
where $\Lambda_{Z^{(\alpha)}}(\theta)$ denotes a single-qubit Pauli channel acting along $\mathcal Z^{(\alpha)}[\cdot] = \hat \sigma_z^{(\alpha)}[\cdot] \hat \sigma_z^{(\alpha)}$ with a random error probability that depends on the stochastic error angle. Twirling thus converts the coherent errors of the bare circuit into probabilistic $Z$ errors.

To understand the global effect of the twirled noise on the circuit, we bring the noise channels associated with each gate layer to the end of the circuit. This leads to a global error channel:
\begin{equation}
   \mathcal U^{(\text{twirled})}_{\text{noisy}}(\vec \theta_1, \dots, \vec \theta_l)\equiv \Lambda_{\text{tot}}(\vec \theta_1, \dots, \vec \theta_l)\mathcal U_{\text{ideal}}, \label{U_twirled_noisy}
\end{equation}
where $\mathcal U^{(\text{twirled})}_{\text{noisy}}(\vec \theta_1, \dots, \vec \theta_l)$ is obtained by replacing all the $Z$-dephasing superoperators in Eq.~\eqref{noisy_prop} by their twirled version via Eq.~\eqref{pauli_layer}:
\begin{equation}
    \mathcal R_{Z}(\vec \theta_j) \rightarrow \prod_{\alpha=1}^{n} \Lambda_{Z^{(\alpha)}}(\theta_j^{(\alpha)}).
\end{equation}
For general circuits obtaining the global error channel $\Lambda_{\text{tot}}(\vec \theta_1, \dots, \vec \theta_l)$ defined above would be an intractable task as it would require commuting the noise through each layer of gates. However, in our case, because Clifford gates merely permute Pauli operators under conjugation, all the twirled noise layers can be commuted through the circuit and grouped at the end. Specifically, the effect of commuting the channel $\Lambda_{Z^{(\alpha)}}(\theta_j^{(\alpha)})$ through the successive layers of Clifford gates is to simply change the axis along which the noise acts. The new axis is simply given by:
\begin{equation}
    \hat P_j^{(\alpha)} = \hat G_{l} \dots \hat G_{j+1}\, 
    \hat Z^{(\alpha)}\, 
    \hat G_{j+1}^{-1} \dots \hat G_{l}^{-1}.
\end{equation}
Denoting these new channels by $\Lambda_{P_{j}^{(\alpha)}}(\theta_j^{(\alpha)})$ and applying this to all Pauli channels in the circuit yields a single effective Pauli channel,
\begin{equation}
    \Lambda_{\text{tot}}(\vec{\theta}_1, \dots, \vec{\theta}_l)
    = \prod_{j=1}^{l} \prod_{\alpha=1}^{n}
        \Lambda_{P_j^{(\alpha)}}(\theta_j^{(\alpha)}).
    \label{eq:global_channel}
\end{equation}

The total noise channel $\Lambda_{\text{tot}}(\vec{\theta}_1, \dots, \vec{\theta}_l)$ we have obtained is a Pauli channel,  which is thus by definition diagonal in the Pauli basis. Obtaining its eigenvalues is therefore sufficient to fully characterize it. 
Note that each channel $\Lambda_{P_j^{(\alpha)}}(\theta_j^{(\alpha)})$ 
in Eq.~(\ref{eq:global_channel}) (i.e.~the circuit-propagated noise from qubit $\alpha$ in layer $j$) has a simple eigenvalue structure.  Eigenvalues are either $1$ (associated with multi-qubit Pauli operators that commute with $\hat{P}_j^{(\alpha)}$) or are $\cos(2\theta_j^{(\alpha)})$ (for multi-qubit Pauli operators that do not commute with
$\hat{P}_j^{(\alpha)}$).  Using this,   
the eigenvalue of the total channel associated with a given multi-qubit Pauli operator $\hat{Q}$ is given by
\begin{equation}
    \lambda_{\hat Q}(\vec{\theta}_1, \dots, \vec{\theta}_l)
    = \prod_{\substack{j,\alpha\\ [\hat{Q}, \hat{P}_j^{(\alpha)}] \neq 0}}
        \cos(2\theta_j^{(\alpha)}).
    \label{eq:total_eigvals}
\end{equation}
This expression directly encodes how the different layers of noise combine to produce the total effective decay of each Pauli component.

In the weak-noise regime of interest
(i.e.~$\lvert \theta_j^{(\alpha)} \rvert \ll 1$), the effect of a single error will be extremely weak, but multiple errors can nonetheless accumulate and severely degrade fidelity (especially if there are noise correlations).  We can capture the physics of this regime by 
approximating each factor in Eq.~(\ref{eq:total_eigvals}) as:
\begin{equation}
    \cos(2\theta) \approx \exp(-2\theta^2).
    \label{eq:cosineApprox}
\end{equation}
This is of course exact to order $\theta^2$. This mild approximation still accurately captures the non-perturbative accumulation of errors across different layers.
The terms dropped here can also accumulate, but such terms are always smaller than the terms we retain (see App.~\ref{app:weak_noise_approx}).  One can thus formally view this as a kind of partial resummation of noise contributions.  Crucially, this approximation does not assume anything about the total error probability of the entire circuit or the correlation time of the noise.

To write Eq.~\eqref{eq:total_eigvals} compactly, we define a $l \times l$ diagonal matrix $\boldsymbol{M}_{\hat Q}^{(\alpha)}$ for each noise operator $\hat Z^{(\alpha)}$ and each Pauli operator $\hat Q$ which encodes whether or not $\hat Q$ commutes with the noise at different time steps,
\begin{equation}
     (\mathbf{M}_{\hat Q}^{(\alpha)})_{jj} =
         \begin{cases}
            1, & [\hat{Q}, \hat{P}_j^{(\alpha)}] \neq 0,\\
            0, & \text{otherwise}.
         \end{cases}
\end{equation}
Using this notation and defining $\mathbf{M}_{\hat Q} = \bigoplus_{\alpha} \mathbf{M}_{\hat Q}^{(\alpha)}$ and $\vec{\theta} = \bigoplus_{\alpha} \vec{\theta}^{(\alpha)}$ with $\vec{\theta}^{(\alpha)} = (\theta_1^{(\alpha)}, \dots, \theta_l^{(\alpha)})$, we can approximately rewrite the eigenvalues of the total channel as the bilinear form:
\begin{equation}
    \lambda_{\hat Q}(\vec{\theta}_1, \dots, \vec{\theta}_l)
    \approx \exp\!\big[-2\, \vec{\theta}^{T} \mathbf{M}_{\hat Q} \vec{\theta}\big] \label{approx_eigvals},
\end{equation}
where the approximate sign comes from replacing the cosines with exponentials. We now have a direct means for characterizing the quality of the entire circuit (for a fixed noise realization, and for a fixed choice of Clifford gates). The total probability that no error occurs during the entire circuit -- which corresponds to the process fidelity \cite{chowUniversalQuantumGate2012} -- can then be expressed as an average over all Pauli eigenvalues \cite{flammia_efficient_2020},
\begin{equation}
    p_{I,\text{tot}}(\vec{\theta}_1, \dots, \vec{\theta}_l)
    = \frac{1}{4^{n}} \sum_{Q} \lambda_{\hat Q}(\vec{\theta}_1, \dots, \vec{\theta}_l).
    \label{eq:prob_no_error}
\end{equation}
While each term in this sum is simple to compute, the total number of terms scales exponentially with $n$, making the computation of $p_{I,\text{tot}}$ intractable in practice.  
Nevertheless, this expression is particularly useful because it leads to a closed-form expression for the circuit fidelity of the entire circuit, which we will use in what follows to gain insight on the effect of noise correlations in the system.

\subsubsection{Averaging over correlated noise}

The next step is to average over the noise variables $\theta_j^{(\alpha)}$. We take these to be zero-mean Gaussian random variables with a covariance matrix $\boldsymbol{\Sigma}$:
% . Its covariance matrix is then simply given by the average (denoted by the overline) of the outer-product of the error angle vector with itself:
\begin{equation}
    \boldsymbol{\Sigma} = \overline{\vec{\theta}\, \vec{\theta}^{T}},\label{covariance_matrix}
\end{equation}  
where overline is used to denote a noise average.  Spatial and temporal noise correlations are thus encoded in the form of $\boldsymbol{\Sigma}$.

For this noise model, the noise average of each eigenvalue $\lambda_{\hat Q}$ of the total noise channel in Eq.~\eqref{approx_eigvals} can be computed exactly (as it is the exponential of a quadratic form), yielding:
\begin{equation}
    \overline{\lambda_{\hat Q}}
    = \frac{1}{\sqrt{\det(\boldsymbol{1} + 4 \mathbf{M}_{\hat Q} \boldsymbol{\Sigma})}}. \label{eigvals}
\end{equation}
It follows that the noise-average of the no-error probability of entire circuit (c.f.~Eq.~\eqref{eq:prob_no_error}) is then simply obtained by summing over all multi-qubit Pauli operators $\hat Q$:
\begin{equation}
    \overline{p_{I,\text{tot}}}
    = \frac{1}{4^{n}} \sum_{\hat Q}
        \frac{1}{\sqrt{\det(\boldsymbol{1} + 4 \mathbf{M}_{\hat Q} \boldsymbol{\Sigma})}}. \label{prob_no_error}
\end{equation}
Finally, the corresponding circuit fidelity as a function of the probability of having no errors is simply given by: \cite{nielsenSimpleFormulaAverage2002}:
\begin{equation}
    F_{\text{tot}} = 
    \frac{2^{n} \overline{p_{I,\text{tot}}} + 1}{2^{n} + 1}.
\end{equation} 
Eq.~(\ref{prob_no_error}) is a central result of this work: it provides a concrete starting point for analytically understanding how different kinds of correlated noise models impact different kinds of Clifford circuit.  The structure of the circuit is encoded in the matrices $\mathbf{M}_{\hat Q}$, while the properties of the noise (including possible spatial and temporal correlations) are encoded in the covariance matrix $\boldsymbol{\Sigma}$.

\subsection{Lower and upper bounds to the circuit fidelity}
\label{subsec:AGF-bounds}

We now use the general analytic results obtained in the previous section to understand how noise correlations impact circuit fidelity after Pauli-twirling.  We start in this section by deriving rigorous bounds on the impact of noise correlations.  Given a particular Clifford circuit and a fixed single gate error rate, we derive bounds on the minimum and maximum possible circuit fidelity.  These correspond to choices for the noise covariance matrix $\boldsymbol{\Sigma}$ that either minimize or maximize correlations (both in space and time).  In particular, the limit of uncorrelated noise yields the worst possible fidelity.

\subsubsection{Pauli-twirling leverages noise correlations}

We start our analysis by writing the diagonals of the noise covariance matrix as:
\begin{equation}
    \operatorname{diag}(\boldsymbol{\Sigma}) 
= ([\sigma_1^{(1)}]^2, [\sigma_2^{(1)}]^2, \ldots, [\sigma_l^{(n)}]^2),
\end{equation}
so that each $[\sigma_j^{(\alpha)}]^2$ characterizes the variance of the error associated with gate layer $j$ and qubit $\alpha$. The matrix $\operatorname{diag}(\boldsymbol{\Sigma})$ therefore contains the information to characterize the noise on each qubit and during each gate in isolation. To isolate the effect that correlations have on the circuit fidelity, we will hold these single qubit and single layer noise variances fixed. For a fixed choice of Clifford circuit and $\{\sigma_j^{(\alpha)}\}$, 
the circuit fidelity can only vary with the off-diagonal elements of $\boldsymbol{\Sigma}$, i.e.~the noise correlations.  We can identify choices of covariance matrices that achieve either a minimum or maximum circuit fidelity.
% %of diagonal elements, 
% one can identify the covariance matrices that either minimize or maximize the circuit fidelity. 
We find that the lower bound to the circuit fidelity (i.e.~worst possible case) is obtained by removing all the correlations between different errors, corresponding to the diagonal covariance matrix 
\begin{equation}
\boldsymbol{\Sigma}_{\text{min}} 
= \operatorname{diag}([\sigma_1^{(1)}]^2, [\sigma_2^{(1)}]^2, \ldots, [\sigma_l^{(n)}]^2).
\end{equation}
This corresponds to the limit where there are neither spatial nor temporal correlations in the noise between different qubits or time steps.

Conversely, we find that the circuit fidelity is maximized when the errors are perfectly correlated (or anti-correlated), i.e., when the covariance matrix is given by an outer product of the form  
\begin{equation}
\boldsymbol{\Sigma}_{\max} = \boldsymbol{\sigma} \, \boldsymbol{\sigma}^T,
\quad \text{with} \quad
\boldsymbol{\sigma} = (\pm\sigma_1^{(1)}, \pm\sigma_2^{(1)}, \ldots, \pm \sigma_l^{(n)})^T. \label{sigma_max}
\end{equation}
At a heuristic level, this form of covariance matrix implies that all the error angles are proportional to a single stochastic variable (hence giving maximal correlations).  
The upshot is that for any set of fixed diagonal elements, the covariance matrices $\boldsymbol{\Sigma}_{\min}$ and $\boldsymbol{\Sigma}_{\max}$ respectively determine the lower and upper bounds on the achievable circuit fidelity. 

To prove the above bounds, we make use of two well-known inequalities for determinants. The lower bound follows directly from the Hadamard inequality applied to positive semi-definite matrices which ensures that each term in Eq.~\eqref{prob_no_error} is lower bounded by:
\begin{equation}
    \frac{1}{\sqrt{\det(\boldsymbol{1} + 4 \mathbf{M}_{\hat Q} \boldsymbol{\Sigma})}} \geq \frac{1}{\sqrt{\det(\boldsymbol{1} + 4 \mathbf{M}_{\hat Q} \boldsymbol{\Sigma}_{\min})}} \label{AGF_bound}.
\end{equation}

On the other hand, the upper bound is obtained by first noting that the matrix $\boldsymbol{M}_{\hat Q} \boldsymbol{\Sigma}$ corresponds to setting to $0$ the rows of $\boldsymbol{\Sigma}$ that correspond to the diagonal entries of $\boldsymbol{M}_{\hat Q}$ which are $0$. The determinant to compute therefore corresponds to $\det(\boldsymbol{1} + 4 \tilde{\boldsymbol{\Sigma}})$ where $\tilde{\boldsymbol{\Sigma}}$ is the principal sub-matrix of $\boldsymbol{\Sigma}$ obtained by removing the rows and columns that correspond to the $0$ diagonal entries of $\boldsymbol{M}_{\hat Q}$. We can in turn bound this determinant from below:
\begin{equation}
    \det(\boldsymbol{1} + 4 \tilde{\boldsymbol{\Sigma}}) \geq 1 + 4 \mathrm{tr}(\tilde{\boldsymbol{\Sigma}}).
\end{equation}
This inequality is a direct consequence of the positive semi-definiteness of $\tilde{\boldsymbol{\Sigma}}$. This bound is saturated by $\boldsymbol{\Sigma_{\text{max}}}$ from Eq.~\eqref{sigma_max} as it is a rank $1$ covariance matrix with eigenvalue $\mathrm{tr}({\boldsymbol{\Sigma}})$ which implies that the similarly defined $\tilde{\boldsymbol{\Sigma}}_{\text{max}}$ is also rank $1$, with eigenvalue $\mathrm{tr}(\tilde{\boldsymbol{\Sigma}}_{\text{max}})$.

This inequality directly implies that the fidelity of a twirled circuit lies between that of the uncorrelated and maximally correlated cases:
\begin{equation}
    F_{\text{tot}}^{\text{(no corr)}} \leq F_{\text{tot}} \leq  F_{\text{tot}}^{(\text{max corr})},
\end{equation}
where $F_{\text{tot}}^{\text{(no corr)}}$ and $F_{\text{tot}}^{(\text{max corr})}$ denote the circuit fidelity associated with $\boldsymbol{\Sigma}_{\text{min}}$ and $\boldsymbol{\Sigma}_{\text{max}}$ respectively.

To summarize, our key finding here is somewhat counterintuitive:  after twirling, correlated Gaussian noise is less harmful than the corresponding uncorrelated version of the noise.  We note that this result goes beyond the findings of Ref.~\cite{winickConceptsConditionsError2022}, which demonstrated that RC effectively tailors coherent noise into stochastic noise, prohibiting a coherent accumulation of errors.  Our result both confirms this result (i.e.~twirling prohibits errors from accumulating quadratically with depth), but also demonstrates that twirling can actually leverage noise correlation to {\it increase} fidelity compared to the uncorrelated case.

We note that while the fidelity of twirled circuits increases with noise correlations, the fact that it is upper bounded also means that the noise cannot be perfectly canceled by twirling. This means that if a particular circuit had been designed to perfectly suppress the noise correlations, such as an optimal dynamical decoupling sequence \cite{violaDynamicalDecouplingOpen1999}, Pauli-twirling would deteriorate the benefits of this decoupling.  However, circuits capable of performing such near-perfect dynamical decoupling are highly atypical.  As we show in Sec.~\ref{subsec:near-term-circuits}, for many generic circuits (including computationally relevant non-Clifford circuits), the effect of twirling is strikingly positive.

\subsubsection{Limiting cases}

To understand the origin of the lower and upper bounds derived in the previous section, we analyze the two extremal covariance matrices, namely the ones corresponding to the limits of Markovian noise with no spatial correlations and of quasistatic noise with maximal spatial correlations.

In the absence of correlations, the correlation matrix is simply diagonal. Here, assuming that each entry is the same (e.g assuming that the noise is stationary) $\boldsymbol{\Sigma}_{jj} = \sigma^2$, we see that each eigenvalue in Eq.~\eqref{eigvals} decays roughly exponentially with the depth of the circuit (provided that the circuit is such that the number of non-zero entries of $\boldsymbol{M}_{\hat Q}$ grows linearly with the depth):
\begin{equation}
    \overline{\lambda_{\hat Q}}^{\mathrm{(Markov)}} = \frac{1}{(1 + 4 \sigma^2)^{\mathrm{tr}(\boldsymbol{M}_{\hat Q})/2}}.
\end{equation}
On the other hand, in the quasistatic limit with maximal spatial correlations, where the covariance matrix elements are $\overline{\theta_j^{(\alpha)} \theta_{j'}^{(\alpha')}} = \sigma^2$, the decay of each eigenvalue follows a power law:
\begin{equation}
    \overline{\lambda_{\hat Q}}^{\mathrm{(qs)}} = \frac{1}{\sqrt{1 + 4 \sigma^2 \mathrm{tr}(\boldsymbol{M}_{\hat Q})}}.
\end{equation}
We therefore, see that in this case correlations dramatically increase the circuit fidelity, going from an exponential decay to a power law decay.

\subsection{Suppression of noise correlations}
\label{subsec:correlation-suppression}

We have just seen that RC can effectively leverage temporal noise correlations to enhance the fidelity. To gain a more concrete understanding of this effect, we analyze how correlations between the error probabilities corresponding to different gate layers are affected by RC, to lowest order in the noise strength. The probability of having no errors in circuit, $p_{I,\text{tot}}$, is given by the sum of the eigenvalues of the global channel [see Eq.~\eqref{prob_no_error}], we therefore study the effect of noise correlations on the decay of the eigenvalues. To do so, we note that $\mathrm{det}(\boldsymbol{1} + 4\boldsymbol{M}_{\hat Q} \boldsymbol{\Sigma})^{-1/2} = \exp(-\mathrm{tr}(\ln (\boldsymbol{1} + 4\boldsymbol{M}_{\hat Q} \boldsymbol{\Sigma}))/2)$ and expand around $\boldsymbol{\Sigma}=0$ to get:
\begin{align}
    \overline{\lambda_{\hat Q}} \approx  1 - 2 \mathrm{tr}[  \boldsymbol{M}_{\hat Q} \boldsymbol{\Sigma}] + 2 \mathrm{tr}[ \boldsymbol{M}_{\hat Q} \boldsymbol{\Sigma}]^2 + 4 \mathrm{tr}[(\boldsymbol{M}_{\hat Q} \boldsymbol{\Sigma})^2]. \label{lambda_expansion}
\end{align}
Remembering that $\boldsymbol{M}_{\hat Q}$ is a diagonal matrix we can easily compute $\mathrm{tr}( \boldsymbol{M}_{\hat Q} \boldsymbol{\Sigma}) =  \sum_{j}(\boldsymbol{M}_{\hat Q})_{jj} \boldsymbol{\Sigma}_{jj}$. Since the diagonal elements of the covariance matrix $\boldsymbol{\Sigma}_{jj}$ correspond to the noise variance for individual qubits and gate layers, the first three terms on the RHS of Eq.~\eqref{lambda_expansion} do not capture the effect of noise correlations across multiple Clifford gate layers. The effect of such correlations therefore only appears to fourth order in the noise strength, through $\mathrm{tr}[( \boldsymbol{M}_{\hat Q} \boldsymbol{\Sigma})^2]$.

In contrast, when the noise is not twirled, the effect of noise correlations on the probability of having no errors generically appears to second order in the noise strength. Since for generic circuits there is no simple formula like Eq.~\eqref{eigvals} for the decay of the eigenvalues, we analyze a simple but illustrative example. We consider a single qubit that is subject to free induction decay. The circuit implementing this would simply correspond to choosing each Clifford gate layer to be the identity. In this case, the probability of having no errors after $l$ layers have been applied is
\begin{equation}
    p_{I,\text{tot}}^{\text{bare}} = \frac{1}{2} + \frac{1}{2} \exp(- \frac{1}{2} \sum_{j,k = 1}^{l} \boldsymbol{\Sigma}_{j,k}),
\end{equation}
which is dependent on noise correlations even to second order in the noise strength.

Having established that twirling suppresses the impact of noise correlations to leading order in the noise strength, we now analyze the residual correlations that survive this suppression. From Eq.~\eqref{eq:twirled_layer}, the error probability on qubit $\alpha$ at layer $j$ is given by $p_j^{(\alpha)} = \sin^2(\theta_j^{(\alpha)})$. A straightforward calculation shows that to 4th order in the noise strength, the covariance between these probabilities for different gate layers, $p_j^{(\alpha)}$ and $p_{j'}^{(\alpha)}$, is
\begin{equation}
\overline{p_j^{(\alpha)} p_{j'}^{(\alpha')}} - \overline{p_j^{(\alpha)}} \ \  \overline{p_{j'}^{(\alpha')}}\approx \frac{1}{2} (\boldsymbol{\Sigma}_{j,j'}^{(\alpha, \alpha')})^2 . 
\end{equation}
Where $\boldsymbol{\Sigma}_{j,j'}^{(\alpha, \alpha')} = \overline{\theta_j^{(\alpha)} \theta_{j'}^{(\alpha')}}$ is the covariance matrix element between the corresponding qubits and gate layers. As expected, the effect of the correlations only appears to fourth order in the noise strength. To gain more intuition, we can analyze a specific type of noise correlations, decaying exponentially in time with correlation time $\tau_c$  and where the strength of the noise is parametrized by $\sigma$:
\begin{equation}
    \boldsymbol{\Sigma}_{jj'}^{(\alpha, \alpha')} = \sigma^2 \delta_{\alpha,\alpha'}\exp(-\frac{\lvert j-j'\rvert t_g}{\tau_c}),
\end{equation}
where $t_g$ is the time required to apply a layer of Clifford gates. In this case, that the correlation between error probabilities is given by:
\begin{equation}
    \overline{p_j^{(\alpha)} p_{j'}^{(\alpha')}} - \overline{p_j^{(\alpha)}} \ \  \overline{p_{j'}^{(\alpha')}} \approx \sigma^4 \delta_{\alpha,\alpha'}\exp(-\frac{\lvert j-j'\rvert t_g}{\tau_c/2}).
\end{equation}
We see that the correlations still decay exponentially with a strength that has been reduced, as expected. Interestingly, the characteristic decay time of the correlations is half that of the physical noise model, suggesting that in this case, twirling not only reduces the strength of the correlations, but also reduces their temporal range.

\section{Extension to more complex settings: finite duration gates and quantum noise}
\label{sec:quantum-finite-gates}

Our results so far focused on a noise model corresponding to dephasing noise along a single axis; this choice provided a tractable setting to analytically describe the effects of Gaussian noise correlations.  

In this section, we show that our results can be generalized to more complicated scenarios.  The first extension is to include the effects of finite-time gate implementations, as well as noise along arbitrary axes. The second extension is to discuss how our approach can be extended to models with quantum noise.  In particular, we show that Pauli-twirling significantly reduces the impact of quantum noise correlations, suggesting that our classical treatment of the noise is representative of more general noise models.

\subsection{Finite duration gates and arbitrary noise axes}
\label{subsec:finite-gates}

We consider here a generalized model in which each Clifford gate in our circuit is applied over a finite time interval.  The full circuit is generated by continuous evolution under a time-dependent Hamiltonian $\hat H_{\text{ctrl},\vec \beta}(t)$, whose form depends on the chosen circuit as well as the sequence of random Pauli-twirling operators $\vec \beta$.  We also let each system qubit $\alpha$ couple to classical stochastic noise sources $\eta^{(\alpha)}(t)$ through the system operators $\hat A^{(\alpha)}$. The noise is again taken to be Gaussian and stationary, and is, hence, fully determined by the two-point correlators $S^{(
\alpha,\alpha'
)}(t-t') = \overline{\eta^{(\alpha)}(t) \eta^{(\alpha')}(t')}$. The total time-dependent Hamiltonian for a given realization of the noise and choice of inserted Pauli gates is then:
\begin{equation}
    \hat H_{\vec \beta}(t) = \hat H_{\text{ctrl}, \vec \beta}(t) + \sum_{\alpha}\eta^{(\alpha)}(t) \hat A^{(\alpha)}. \label{classical_H_tot}
\end{equation}
% This way of modeling the noisy implementation of the gates is strictly more general than the Z-axis phase errors considered in the previous section. 
To recover the simpler model of the previous sections, one can take the limit of vanishing gate durations and assume that noise acts along the $Z$ axis during an idle period following each gate.

To make this more general setting tractable, we consider the common scenario 
of noise with correlation time much greater than the Clifford gates duration. We can then safely replace the time-dependent noise $\eta^{(\alpha)}(t)$ during the time interval of a specific gate by its time-average during that interval: $\theta_j^{(\alpha)} \equiv \int_{jt_G}^{(j+1)t_G}dt ~ \eta^{(\alpha)}(t)$. 
Note that the effects of finite-duration gates persist under this approximation, as we show below. 

We proceed in an analogous manner to our treatment of instantaneous gates. We sketch the approach here, providing full details in App.~\ref{app:finite-duration}. First, we average the noise channels during each gate layer over the random Pauli gates. As a result, noise at each timestep is described by a stochastic Pauli channel. Second, we commute all Pauli channels to the end of the circuit, to obtain an effective, total noise channel, as in Eq.~\eqref{U_twirled_noisy}. Finally, we find the eigenvalues of this total noise channel for a given realization of the random phase errors $\vec \theta$. Surprisingly, despite the additional complexity of having finite-time gates and arbitrary noise axes, the eigenvalues can still be expressed as the exponential of a quadratic form in the coarse-grained noise variables $\vec \theta$:
\begin{equation}
    \lambda_{\hat Q}^{(\text{f-t})}(\vec{\theta}_1, \dots, \vec{\theta}_l)
    \approx \exp\!\big[-2\, \vec{\theta}^{T} \mathbf{M}^{(\text{f-t)}}_{\hat Q} \vec{\theta}\big], \label{finite_duration_M}
\end{equation}
where the superscript ``(f-t)'' denotes the finite-time generalization of the matrix $\mathbf{M}_{\hat Q}$ introduced for instantaneous gates. This expression is the direct analogue of  Eq.~\eqref{approx_eigvals}, with some crucial differences. Instead of being diagonal, the matrix $\mathbf{M}^{(\text{f-t)}}_{\hat Q}$ now has block structure:
\begin{equation}
    \mathbf{M}^{(\text{f-t)}}_{\hat Q}
    =
    \begin{pmatrix}
        (\boldsymbol{M}^{\text{(f-t)}}_{\hat Q})^{(1,1)} &   \cdots   & (\boldsymbol{M}^{\text{(f-t)}}_{\hat Q})^{(1,n)} \\
        \vdots  & \ddots  & \vdots  \\
        (\boldsymbol{M}^{\text{(f-t)}}_{\hat Q})^{(n,1)} & \cdots  & (\boldsymbol{M}^{\text{(f-t)}}_{\hat Q})^{(n,n)}
    \end{pmatrix},
\end{equation}
where the blocks are indexed by the qubit indices. Each block is diagonal: $(\boldsymbol{M}^{\text{(f-t)}}_{\hat Q})^{(\alpha, \alpha')} = \mathrm{diag}((m_{1,\hat Q}^{\text{(f-t)}})^{(\alpha,\alpha')}, \dots, (m_{l,\hat Q}^{\text{(f-t)}})^{(\alpha,\alpha')})$, with entries $(m_{l,\hat Q}^{\text{(f-t)}})^{(\alpha,\alpha')} \in \mathbb R$. The fact that the values are not simply zeroes and ones is a consequence of the fact that during a finite-time Clifford gate layer, the noise operator is time-dependent and will generically be a linear combination of multiple Pauli operators. Furthermore, the fact that there are off-diagonal blocks (with $\alpha \neq \alpha'$) indicates that spatial correlations can enter in leading order in the noise strength. This effect is due to the fact that two initially independent noise operators acting on different qubits can evolve to have a non-zero overlap during a single Clifford gate layer. This is in stark contrast to the case of instantaneous gates, where noise operators remain orthogonal at all times.

In App.~\ref{app:finite-duration}, we show how $\mathbf{M}^{(\text{f-t})}_{\hat Q}$ can be computed. This procedure unfortunately becomes infeasible for large circuits as it requires computing the time-evolution of each noise operator for each Clifford gate layer. While the above result then does not provide a useful starting point for large-scale simulation, it still has utility in providing qualitative understanding of noise correlations and the impact of finite-duration gates. 

Crucially, even in this more complex setting, we can obtain bounds on the circuit fidelity that are analogous to Eq.~\eqref{AGF_bound}. As shown in App.~\ref{app:finite-duration}, the upper bound is the same as in the previous section, corresponding to the maximally correlated covariance matrix Eq.~\eqref{sigma_max}. On the other hand, to derive the lower bound, we consider a slightly modified setting. In addition to fixing the noise variance for each qubit during each gate layer, we also fix the covariances corresponding to spatial correlations between different qubits during a given gate layer. These correspond to the covariance matrix elements which are diagonal in the layer index, $\{\boldsymbol{\Sigma}_{j,j}^{(\alpha,\alpha')}\}_{j,\alpha,\alpha'}$. Given these fixed elements, we ask which covariance matrix minimizes the circuit fidelity. In App.~\ref{app:finite-duration}, we show that the minimum is obtained when all off-diagonal elements in the layer index vanish, i.e.,
\begin{equation}
\boldsymbol{\Sigma}_{j,j'}^{(\alpha,\alpha')} = 0 \quad \text{for } j \neq j'.
\end{equation}
This lower bound implies that temporal correlations in the noise generically increase the circuit fidelity.

\subsection{Reduction of quantum correlations}
\label{subsec:quantum-noise}

We next ask how our main results are modified by allowing the noise driving our qubits to be quantum noise, i.e.~stemming from a coupling to a quantum environment.  We will show that Pauli-twirling highly reduces the effect of the quantum part of the temporal correlations. Specifically, we show that to second order in the noise strength, quantum correlations are completely suppressed by twirling, meaning that they can only matter for long correlation times or very strong noise. 

To extend our setup to include a quantum bath, we take a very similar starting point as Eq.~\eqref{classical_H_tot} but replace the classical noise processes $\eta^{(\alpha)}(t)$ by time-dependent bath operators $\hat B^{(\alpha)}(t)$ that evolve according to a bath-only Hamiltonian $\hat H_B$ \footnote{This means that our starting point is an interaction picture with respect to the bath only Hamiltonian, defined by $\hat U_B(t) = e^{- i \hat H_B t}$.}. We assume that the operator valued process $\hat B^{(\alpha)}(t)$ is Gaussian and stationary with respect to the initial state of the bath $\hat \rho_B$, meaning that the effect of the bath on the system is fully characterized by the two-point correlators $S^{(\alpha,\alpha')}(t_1-t_2) = \mathrm{tr}[\hat B^{(\alpha)}(t_1) \hat B^{(\alpha')}(t_2) \hat \rho_B]$. Note that because the bath operators do not commute with one another at all times, $S^{(\alpha,\alpha')}(t)$ will generally be a complex number with real and imaginary part:
\begin{equation}
    S^{(\alpha,\alpha')}(t) = S_R^{(\alpha,\alpha')}(t) + i \epsilon S_I^{(\alpha,\alpha')}(t),
\end{equation}
where $\epsilon$ is a dimensionless bookkeeping parameter. The two components are obtained from the symmetric (anti-commutator) and antisymmetric (commutator) parts of the operator correlator:
\begin{align}
    S_R^{(\alpha,\alpha')}(t)
    &= \frac{1}{2}\,
       \mathrm{tr}\!\left(
           \left\{
               \hat B^{(\alpha)}(t),
               \hat B^{(\alpha')}(0)
           \right\}
           \hat \rho_B
       \right), 
       \label{SR_def}
    \\
    S_I^{(\alpha,\alpha')}(t)
    &= \frac{1}{2i}\,
       \mathrm{tr}\!\left(
           \left[
               \hat B^{(\alpha)}(t),
               \hat B^{(\alpha')}(0)
           \right]
           \hat \rho_B
       \right).
       \label{SI_def}
\end{align}

For times $t \geq 0$, the imaginary part of the two-point correlator is directly related to the bath susceptibilities. Using Eqs.~\eqref{SR_def}–\eqref{SI_def}, one finds
\begin{equation}
\chi^{(\alpha,\alpha')}(t-t')
= 2\, \theta(t-t')\,
S_I^{(\alpha,\alpha')}(t-t').
\label{chi_SI_relation}
\end{equation}
These are the standard Kubo linear response functions: they characterize the change in the expectation value of $\hat B^{(\alpha)}(t)$ due to a weak perturbation of the bath Hamiltonian through $\hat B^{(\alpha')}(t')$ at earlier time \cite{clerkIntroductionQuantumNoise2010}. 

The real part, $S_R^{(\alpha,\alpha')}(t)$, gives rise to the same classical noise physics that we considered in the previous sections. The classical noise limit can be achieved by taking a high temperature bath. In this case $S^{(\alpha,\alpha')}_R(t)$ is proportional to the temperature while $S^{(\alpha,\alpha')}_I(t)$ is constant. As a result, the effect of $S^{(\alpha,\alpha')}_I(t)$ becomes negligible in comparison to $S^{(\alpha,\alpha')}_R(t)$.

As in Eq.~\eqref{classical_H_tot}, the system evolves according to a time-dependent Hamiltonian, $\hat H_{\vec \beta}(t)$, implementing the Clifford circuit as well as the random Pauli gates which are indexed by $\vec \beta$. To make progress, we move to the interaction frame with respect to this Hamiltonian, defined by $\hat U_{\text{ctrl}, \vec \beta}(t) = \mathcal T\exp(-i \int_0^tdt' \hat H_{\text{ctrl}, \vec \beta}(t'))$. In this frame, the Hamiltonian is:
\begin{equation}
    \hat H_{\vec \beta}(t) =  \sum_{\alpha} \hat A^{(\alpha)}_{\vec \beta}(t) \otimes \hat B^{(\alpha)}(t),
\end{equation}
where $\hat A^{(\alpha)}_{\vec \beta}(t) = \hat U_{\text{ctrl}, \vec \beta}^{\dagger}(t) \hat A^{(\alpha)}\hat U_{\text{ctrl}, \vec \beta}(t)$ are the time-evolved noise operators. 

Once again, we want to understand the dynamics averaged both over the noise and the Pauli gates. Unlike in previous sections, however, we average over the noise first. A common approach to obtain the resulting reduced dynamics is to derive a time-convolutionless master equation \cite{lidar_lecture_2020}. To leading order in the system-bath coupling, we obtain:
\begin{equation}
     \dot{\hat{\rho}}_{\vec \beta}(t) =  \int_0^{t} dt_1 \left(\mathcal K_{\vec \beta}^{(\text{cl})}(t,t_1) + \epsilon \mathcal K_{\vec \beta}^{(\text{qu})}(t,t_1) \right) \hat \rho_{\vec \beta}(t), \label{TCL}
\end{equation}
where $\mathcal K_{\vec \beta}^{(\text{cl,qu})}(t_1,t_2)$ are memory kernels defined explicitly in App. \ref{app:quantum-noise}.

To compute the average over the random Pauli gates, we follow a procedure introduced in Ref.~\cite{brillant_randomized_2025} and perform a generalized cumulant expansion in the random gate sequences $\vec \beta$. This approximation corresponds to factoring the averages of the memory kernels $\mathcal K_{\vec \beta}^{\text{cl,qu}}$ and the state $\hat \rho_{\vec \beta}$ when averaging Eq.~\eqref{TCL} over the random Pauli gates. 

Since, we want to understand the effect of twirling on the part of the dynamics due to $\mathcal K_{\vec \beta}^{(\text{qu})}(t_1,t_2)$, the object to average is:
\begin{multline}
    \mathcal K_{\vec \beta}^{(\text{qu})}(t_1,t_2)[\cdot] \\= \frac{1}{2}\sum_{\alpha, \alpha'} \chi^{(\alpha, \alpha')}(t_1-t_2)  [ \hat A^{(\alpha)}_{\vec \beta}(t_1), \{ \hat A^{(\alpha')}_{\vec \beta}(t_2), \cdot \}].\label{quantum_colonel}
\end{multline}

We expand the nested commutator and anti-commutator to obtain:
\begin{widetext}
\begin{equation}
    [ \hat A^{(\alpha)}_{\vec \beta}(t_1), \{ \hat A^{(\alpha')}_{\vec \beta}(t_2), \hat \rho \}] = \hat A^{(\alpha)}_{\vec \beta}(t_1) \hat A^{(\alpha')}_{\vec \beta}(t_2) \hat \rho - \hat \rho \hat A^{(\alpha')}_{\vec \beta}(t_2) \hat A^{(\alpha)}_{\vec \beta}(t_1) + \hat A^{(\alpha)}_{\vec \beta}(t_1) \hat \rho \hat A^{(\alpha')}_{\vec \beta}(t_2) - \hat A^{(\alpha')}_{\vec \beta}(t_2)  \hat \rho \hat A^{(\alpha)}_{\vec \beta}(t_1). \label{quantum_kernel_expanded}
\end{equation}
\end{widetext} 
Upon averaging over the Pauli gates indexed by $\vec \beta$, the right hand side of this expression vanishes exactly, meaning that to lowest order in the system-bath coupling strength, the quantum part of the noise is suppressed. We provide a simple heuristic for this result here and leave details of the exact calculation to App.~\ref{app:quantum-noise}.

To build intuition for that result, suppose that the system operator at earlier time $t_2$ approximately commutes with the state of the system and with other system operators at later time $t_1$:
\begin{equation}
    \left[ \hat A^{(\alpha')}_{\vec\beta}(t_2), \hat\rho \right] \approx 0 \quad \text{and}  \quad \left[ \hat A^{(\alpha')}_{\vec\beta}(t_2), \hat A^{(\alpha)}_{\vec\beta}(t_1) \right] \approx 0,\label{commuting_approx}
\end{equation}
which is analogous to replacing $\hat A^{(\alpha')}_{\vec\beta}(t_2)$ by its expectation value. In this case, we could approximate Eq.~\eqref{quantum_kernel_expanded} by:
\begin{equation}
    \left[ \hat A^{(\alpha)}_{\vec\beta}(t_1),
    \left\{ \hat A^{(\alpha')}_{\vec\beta}(t_2), \hat\rho \right\} \right]
    \approx 
    2 \left[ 
    \hat A^{(\alpha)}_{\vec\beta}(t_1)
    \hat A^{(\alpha')}_{\vec\beta}(t_2),
    \hat\rho
    \right].
\end{equation}
Thus, at this level of approximation, the quantum kernel in our master equation generates unitary evolution under an effective Hamiltonian
\begin{equation}
    \hat H_{\mathrm{eff}}(t)
    = \frac{1}{2}\sum_{\alpha, \alpha'} \int_0^t dt' \chi^{(\alpha, \alpha')}(t-t')
    \hat A^{(\alpha)}_{\vec\beta}(t)
    \hat A^{(\alpha')}_{\vec\beta}(t').
\end{equation}
This describes a bath-mediated qubit-qubit interaction, and directly involves the linear response susceptibility of the bath.  
The random Pauli gates indexed by $\vec\beta$ randomize the axis and sign of this effective interaction. Averaging over $\vec\beta$ therefore removes all traceless components of the effective coupling. Any remaining identity contribution commutes with $\hat\rho$ and hence drops out of the commutator. Consequently, the quantum kernel, Eq.~\eqref{quantum_colonel},  vanishes to lowest order in the system-bath coupling strength.

While the approximation in Eq.~(\ref{commuting_approx}) provides useful intuition, we stress that the leading-order cancellation of quantum effects holds even without this approximation: each term in the full expression Eq.~\eqref{quantum_colonel} vanishes upon averaging over the random Pauli gates (see App.~\ref{app:quantum-noise}).

\section{Extensions to non-Clifford circuits}
\label{sec:applications}

Our previous analysis of correlated noise in Pauli-twirled circuits focused exclusively on Clifford circuits, a restriction that enabled us to obtain analytic insights and rigorous bounds.  In this section, we explore how this general understanding also applies to settings that go beyond Clifford circuits.  While an analytic treatment is no longer feasible, we present both heuristic arguments and explicit numerical evidence suggesting that at a qualitative level, our general conclusions on the impact of twirling and noise correlations remain valid. In particular, we argue that spatio-temporal noise correlations tend to have a \emph{beneficial} effect on the circuit fidelity beyond the Clifford setting.  Note that in the absence of noise correlations, previous work has suggested that many features of how noise impacts general circuits can be captured using Clifford-circuit approximations \cite{merkel_when_2025}.  Here, we suggest something analogous holds even with noise correlations.   

\subsection{Random circuits and twirling-induced reduction of circuit-to-circuit fidelity variation} \label{sec:variance}

Beyond its impact on the circuit fidelity, Pauli-twirling also strongly reduces the sensitivity of the fidelity to the specific circuit realization. For a fixed noise model and fixed circuit depth, different random circuits typically exhibit noticeable variation in noise-averaged circuit fidelity. As we now show, this circuit-to-circuit variation is strongly suppressed under Pauli-twirling.

The intuition is as follows. For a given logical circuit $C$, Pauli-twirling replaces the fidelity of a single physical implementation by an average over a family of Pauli-dressed implementations $\{ C_{\vec \beta} \}$ that realize the same ideal unitary. The corresponding noise-averaged circuit fidelity in the twirled description can be written as
\begin{equation}
    F_{\mathrm{tw}}(C)
    =
    \langle F(C_{\vec \beta}) \rangle_{\vec \beta},
\end{equation}
where $F(C_{\vec \beta})$ is the circuit fidelity of the bare circuit $C$ for a particular realization of the Pauli gates $\vec \beta$. Thus, twirling replaces the fidelity of one particular compiled implementation by an average over many logically equivalent randomizations. This averaging suppresses implementation-dependent coherent interference effects, thereby reducing circuit-to-circuit fidelity fluctuations.

\begin{figure}
    \centering
    \includegraphics[width=1\linewidth]{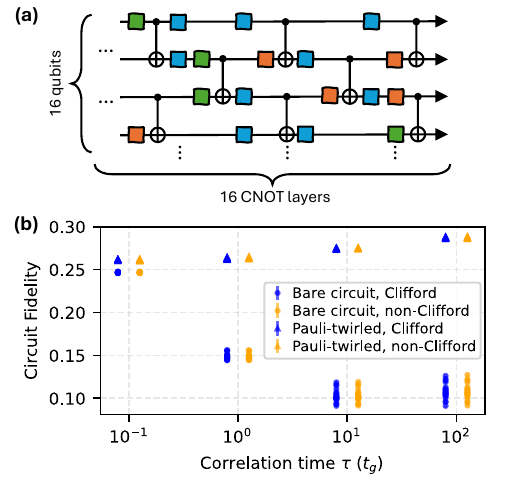}
    \caption{
    Noise-averaged circuit fidelity of random Clifford and non-Clifford brickwork circuits under non-Markovian $Z$-dephasing, comparing the bare noise model to its Pauli-twirled effective description.
    (a) Circuit construction.
    We generate $20$ random depth-$16$ circuits on $16$ qubits composed of alternating layers of nearest-neighbor CNOT gates and single-qubit layers in which a gate is applied independently to each qubit with probability $0.3$, chosen to be $S$ or $\sqrt{X}$ with equal probability.
    Non-Clifford circuits are obtained by additionally inserting a $T$ gate independently for each qubit with probability $0.5$.
    (b) Circuit fidelity as a function of noise correlation time. The dephasing noise has covariance given by Eq.~\eqref{corr_func} with $\sigma=0.15$; data are shown for correlation times $\tau/t_g = 0.1, 1, 10,$ and $100$.
    Error bars from noise averaging are smaller than the marker size, so the visible spread reflects circuit-to-circuit variation.
    Under Pauli-twirling, fidelities are strongly concentrated across circuit instances, whereas in the bare model substantial circuit-to-circuit variation is observed.
    }
    \label{fig:variance_fig}
\end{figure}

Figure~\ref{fig:variance_fig} provides numerical evidence of this effect. We generate $20$ random brickwork circuits on $16$ qubits corresponding to two circuit ensembles:

(A) Clifford brickwork circuits composed of alternating layers of nearest-neighbor CNOT gates and random single-qubit Clifford gates.

(B) Non-Clifford circuits obtained from (A) by additionally inserting random $T$ gates.

Between each CNOT layer, a single-qubit gate is applied independently to each qubit with probability $0.3$, chosen to be $S$ or $\sqrt{X}$ with equal probability. 
In the non-Clifford ensemble, an additional $T$ gate is inserted independently on each qubit with probability $0.5$.

To model the noise, we apply single qubit $Z$-dephasing errors following each CNOT gate. The noise is fully characterized by the covariance matrix:
\begin{equation}
    \overline{\theta_{j}^{(\alpha)} \theta_{j'}^{(\alpha')}} 
    = \delta_{\alpha,\alpha'} \, \sigma^2 e^{-(j-j')t_G/\tau_c}, \label{corr_func}
\end{equation}
where $\sigma$ sets the overall noise strength and $\tau_c$ characterizes the correlation time. We restrict ourselves to temporal correlations only, as these already capture the coherent error accumulation that drives circuit-to-circuit fidelity variation.

For each randomly generated circuit, we compute the noise-averaged circuit fidelity for both the bare noise model described above and its twirled version. In the twirled case, the circuit fidelities are strongly concentrated around their mean for both the Clifford and non-Clifford ensembles. We also notice that they depend only weakly on the correlation time of the noise. This is a consequence of the fact that noise correlations affect the twirled fidelity only at higher orders in the noise strength. Importantly, we note that both Clifford and non-Clifford circuit fidelities increase with the correlation time of the noise, which is consistent with the bounds derived in Sec.~\ref{subsec:AGF-bounds}.

Additionally, for the circuit ensembles considered here, twirling noticeably increases the mean fidelity. We emphasize that this enhancement is not universal. For example, if the single-qubit gates inserted between CNOT layers were drawn uniformly from the Clifford group, the twirled and bare fidelities would have the same mean. In that case, twirling would improve the fidelity for only approximately half of the circuits, as shown in App.~\ref{app:numerical-simulations}.

\subsection{Near-term circuits}
\label{subsec:near-term-circuits}

We now turn to mid-sized quantum circuits containing non-Clifford gates.  
As shown in the previous section, Pauli-twirling strongly suppresses circuit-to-circuit fidelity fluctuations in random ensembles.
While this lowering of the variance is encouraging, one might worry that realistic circuits—often highly structured and algorithm-specific—may deviate substantially from typical random instances.  
To address this concern, we analyze two standard verification circuits for 18-qubit processors provided in QASM \cite{cross_openqasm_2022, cross_open_2017}: (i) the quantum Fourier transform (QFT), and
(ii) a square-root algorithm based on amplitude amplification. For both cases, we implemented the circuits using the transpiled OpenQASM representations (\texttt{qft\_n\_18.qasm} and \texttt{sqrt\_n\_18.qasm}) \cite{cross_openqasm_2022, cross_open_2017}.

Starting from these transpiled circuits, we introduce noise in the same way as in the previous section: after each two-qubit gate, we apply $Z$-axis phase errors to both participating qubits. The covariance matrix of the noise is the same as in the previous section (see Eq.~\eqref{corr_func}). Here, the correlation time refers to the number of two-qubit gates over which phase errors on a given qubit remain correlated, rather than a microscopic physical timescale. We consider two implementations of the noise. The bare coherent model, where $Z$ rotations are applied directly, and its twirled version. In both cases, we average over many noise realizations and initial states and compute the circuit fidelity.

\begin{figure}
    \centering
    \includegraphics[width=\linewidth]{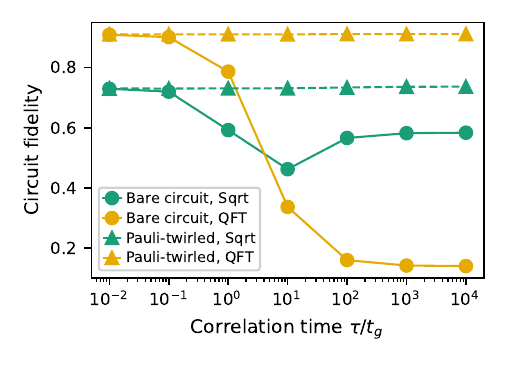}
    \caption{
    Circuit fidelity as a function of the noise correlation time.
    We simulate the circuit fidelity of the \texttt{qft\_n\_18.qasm} (red) and \texttt{sqrt\_n\_18.qasm} (blue) circuits subject to non-Markovian dephasing noise with covariance given by Eq.~\eqref{corr_func} and noise strength $\sigma = 0.035$.
    For each circuit, we compare the bare implementation (circles) to its Pauli-twirled counterpart (triangles). Error bars corresponding to the standard deviation with respect to the noise averaging were calculated but are too small to be noticeable.
    In the Pauli-twirled case, the state fidelity exhibits only a weak dependence on the correlation time, whereas in the bare case it varies strongly as temporal correlations increase.
    }
    \label{fig:near_term_circuits}
\end{figure}

Figure~\ref{fig:near_term_circuits} shows the resulting circuit fidelities as a function of the correlation time $\tau_c$. In the bare model, fidelities decrease substantially as temporal correlations increase, reflecting constructive accumulation of coherent phase errors across successive gates. For the square-root algorithm, the dependence on $\tau_c$ is non-monotonic, indicating that both constructive and destructive interference between coherent errors can occur. In contrast, under Pauli-twirling the fidelity exhibits only a weak and predominantly monotonic dependence on $\tau_c$. Specifically, in the twirled version of the \texttt{sqrt\_n\_18.qasm} circuit, the fidelity goes from approximately $0.730$ for $\tau_c = 10^{-2} t_g$ to $0.736$ for $\tau_c = 10^{4} t_g$. On the other hand, for the \texttt{qft\_n\_18.qasm} circuit, there is no change within our statistical error bars ($0.001$).  Although the enhancement is modest, it is consistent with the lower bound derived in Sec.~\ref{subsec:AGF-bounds} and suggests that temporal correlations can enhance the fidelity of twirled circuits even beyond the Clifford regime.

\subsection{Quantum error correction}
\label{subsec:QEC}

We conclude this section by analyzing another practical setting in which the assumptions underlying our analytical bound no longer strictly apply. Specifically, we consider one of the simplest quantum error-correcting codes: the three-qubit phase-flip repetition code \([[3,1,1]]\).  Although this code involves only Clifford operations, the inclusion of measurement and feedback steps in the error-correction procedure means that the assumptions underlying our earlier derivations no longer strictly apply. 

We encode a single logical qubit using this code and study the survival probability after $250$ rounds of error correction. As before, we apply $Z$-axis phase errors following each two-qubit gate, with no spatial correlations between qubits. We stress that our noise model allows for noise correlations between different error correction cycles.  It is important to note that, while we previously conjectured that the circuit fidelity of non-Clifford circuits increases with noise correlations, the same conclusion does not immediately follow for the logical fidelity in an error-correcting code. Even if the overall physical error rate decreases with increasing correlation time, this does not necessarily imply a reduction in the rate of uncorrectable logical errors.  Correlated error patterns may, in principle, increase logical failure probabilities.

\begin{figure}
    \centering
    \includegraphics[width=\linewidth]{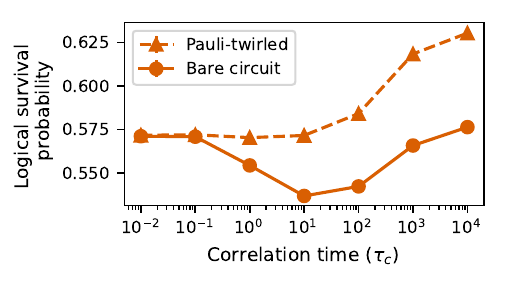}
    \caption{
    Logical survival probability of the \([[3,1,1]]\) repetition code initialized in the logical $\ket 0$ state after $250$ error-correction cycles as a function of the noise correlation time.
    Each data and ancilla qubit is subject to non-Markovian dephasing noise after each two-qubit gate, with correlation function~Eq.\eqref{corr_func} and noise strength $\sigma = 0.05$.
    We compare the Pauli-twirled (green) and bare (orange) circuit implementations.
    Error bars corresponding to the standard deviation with respect to the noise averaging were calculated but are too small to be noticeable. In the Pauli-twirled case, the logical survival probability increases slightly with correlation time, whereas in the bare case it exhibits a pronounced non-monotonic dependence.
    \label{fig:repetition_code}
}
\end{figure}

For the specific QEC setting studied here, we find that Pauli-twirling enhances the logical fidelity over a range of different noise correlation times $\tau_c$.. As shown in Fig.~\ref{fig:repetition_code}, the logical survival probability after 250 error-correction rounds increases monotonically with $\tau_c$ in the twirled model. In contrast, in the bare circuit case the fidelity exhibits a non-monotonic dependence on $\tau_c$.

The behavior of the bare circuit fidelity can be understood by examining the effect of temporally correlated phase errors on the ancilla qubits. When correlations persist primarily within a single error-correction cycle, coherent phase errors can accumulate and propagate through the feedback procedure, reducing the logical fidelity. When correlations extend across multiple cycles, however, successive coherent rotations may partially cancel, mitigating their net impact.
In contrast, for the twirled noise model, such coherent interference effects are removed, and the logical fidelity is governed primarily by the effective stochastic error rate, which depends only weakly on $\tau_c$. Although the overall dependence on $\tau_c$ remains modest, these results indicate that Pauli-twirling mitigates the sensitivity of logical performance to temporal correlations even in error-correction protocols. 

While the above results are for a specific and simple QEC scenario, they suggest that the effects of twirling and noise correlations we established rigorously for non-adaptive Clifford circuits can still qualitatively hold in more complex QEC setting. 

\section{Discussion}
\label{sec:discussion}

We have investigated the effects of noise correlations in Pauli-twirled circuits. Using a generalized cumulant expansion, we derived tight bounds on the average circuit fidelity of twirled Clifford circuits subject to classical, temporally and spatially correlated noise. A key and perhaps counterintuitive result is that, for a fixed single-gate noise variance, the worst-case average circuit fidelity arises for noise that is \emph{neither temporally nor spatially correlated}. In other words, any amount of correlation can be seen as a resource in Pauli-twirled circuits. This challenges the common intuition that non-Markovian and spatially correlated noise is generically more detrimental than uncorrelated noise in quantum circuits.

Our bounds readily extend to more realistic settings, including noise coupling along arbitrary axes and finite-duration gate implementations. In this case, the primary qualitative modification is an enhanced sensitivity to spatial noise correlations accross different qubits.

We also analyzed the effect of Pauli-twirling on genuinely quantum, non-Markovian noise. Using time-convolutionless master equation techniques, we showed that Pauli-twirling suppresses quantum correlations between the system and its environment to lowest order in the system-bath coupling.

Finally, we numerically tested our predictions beyond the Clifford setting by studying both near-term algorithmic circuits and random non-Clifford circuits. In these more general regimes, we find that Pauli-twirling continues to improve the overall circuit fidelity, despite the absence of an exact Clifford structure.

Taken together, these results indicate that bounds derived for Clifford circuits and classical noise remain informative for more general noise models and circuit architectures. More broadly, they highlight that Pauli-twirling can be a powerful and robust tool for mitigating temporally and spatially correlated noise in near-term quantum devices.

\begin{acknowledgments}

 This research was sponsored by the Army Research Office and was 
 accomplished under Grant No. W911NF-23-1-0116. A.B. acknowledges support of the FRQNT through a doctoral scholarship.  A.C. acknowledges support from the Simons Foundation through a Simons Investigator Award (Grant No. 669487). This manuscript has been coauthored by UT-Battelle, LLC, under Contract No. DE-AC05-00OR22725 with the U.S. Department of Energy (DOE). This research used resources of the Oak Ridge Leadership Computing Facility, which is a DOE Office of Science User Facility supported under Contract DE-AC05-00OR22725.

\end{acknowledgments}

\appendix

\section{Weak noise approximation}
\label{app:weak_noise_approx}

In this section, show that the approximation made in Eq.~\eqref{eq:cosineApprox} is justified. Specifically, we show that to any order in $\theta$, the terms that are dropped are negligible compared to the ones we keep to any order in the noise strength.

Starting from Eq.~\eqref{eq:total_eigvals}, we write:
\begin{equation}
    \log(\lambda_{\hat Q}(\vec{\theta}_1, \dots, \vec{\theta}_l))
    = \sum_{\substack{j,\alpha\\ [\hat{Q}, \hat{P}_j^{(\alpha)}] \neq 0}}
       \log\!\left(\cos(2\theta_j^{(\alpha)})\right).
\end{equation}
For $|\theta|\ll 1$, we can expand the logarithm:
\begin{equation}
\log(\cos(2\theta))
=
-2\theta^2 + O(\theta^4).
\end{equation}
Substituting this in the previous expression, we obtain:
\begin{equation}
    \log(\lambda_{\hat Q}) =
    -2 \sum_{\substack{j,\alpha\\ [\hat{Q}, \hat{P}_j^{(\alpha)}] \neq 0}}
       (\theta_j^{(\alpha)})^2 
       +
       O\!\left(
    \sum_{j,\alpha}
    (\theta_j^{(\alpha)})^4
    \right).
\end{equation}
We can then exponentiate both sides, yielding:
\begin{equation}
    \lambda_{\hat Q} =
    \exp(-2 \sum_{\substack{j,\alpha\\ [\hat{Q}, \hat{P}_j^{(\alpha)}] \neq 0}}
       (\theta_j^{(\alpha)})^2 
       +
       O\!\left(
    \sum_{j,\alpha}
    (\theta_j^{(\alpha)})^4
    \right)).
\end{equation}
Neglecting the terms $O\!\left(\sum_{j,\alpha}(\theta_j^{(\alpha)})^4\right)$ in the exponential is therefore equivalent to the approximation made in Eq.~\eqref{eq:cosineApprox}. We can then see that the term that we neglect are smaller than the ones that we keep. Indeed, we have:
\begin{equation}
\sum_{j,\alpha} (\theta_j^{(\alpha)})^4
\le
\Big(
\max_{j,\alpha} (\theta_j^{(\alpha)})^2
\Big)
\sum_{j,\alpha} (\theta_j^{(\alpha)})^2.
\end{equation}

This shows that the relative correction to the exponent is bounded by
$\max_{j,\alpha}(\theta_j^{(\alpha)})^2$, which is small whenever
the noise during each individual gate is weak. Importantly,
this argument places no restriction on the total accumulated
error $\sum_{j,\alpha}(\theta_j^{(\alpha)})^2$ or on the
temporal correlations of the noise. We can therefore see this approximation as a partial resummation of the contributions of the noise to all orders in the noise strength.

\section{Derivation of total noise channel for the finite-duration gates}
\label{app:finite-duration}

In this section we provide the derivations that led to the results presented in Sec.~\ref{subsec:finite-gates}. We consider that a bare Clifford circuit, is implemented in a time $t_G$ via a control Hamiltonian $\hat H_{\text{ctrl}}$ which can be exponentiated to obtain the various gates in the sequence:
\begin{equation}
    \hat g_{j} = \mathcal T\exp(-i\int_{(j-1)t_G}^{jt_G} dt' \hat H_{\text{ctrl}}(t')),
\end{equation}
where $j$ is indexing the Clifford gate being applied between $(j-1)t_G$ and $j t_G$. While the Clifford gates are being applied, the system is subject to classical noise modeled by the time-dependent random variables $\eta^{(\alpha)}(t)$ that each couple to Hermitian system operators $\hat A^{(\alpha)}$. For instance, one could think of $\alpha$ as a qubit index and take $\hat A^{(\alpha)} = \hat \sigma_z^{(\alpha)}$ such that the noise simply leads to classical dephasing along the $Z$ axis of each qubit.  The total dynamics for one realization of the noise is therefore modeled by the Hamiltonian:
\begin{equation}
    \hat H_{\vec \eta}(t) = \hat H_{\text{ctrl}}(t) + \sum_{\alpha} \eta^{(\alpha)}(t) \hat A^{(\alpha)}.\label{noisy_hamiltonian},
\end{equation}
Where the subscript $\vec \eta$ denotes the explicit dependence on the noise. % We consider that the Pauli gates are applied instantaneously and perfectly. This assumption is well motivated by current hardware where single qubit gates are applied much faster than there multi-qubit counterpart.

For a given realization of the noise random variables $\eta^{(\alpha)}(t)$, the noisy implementation of the $j$th gate is:
\begin{equation}
    \tilde g_{j,\vec \eta} = \mathcal T\exp(-i\int_{(j-1)t_G}^{jt_G} dt' \hat H_{\vec \eta}(t')). \label{noisy_unitary_gate}
\end{equation}
To isolate the effect of twirling, it is convenient to separate the contributions of the control Hamiltonian and the noise. Passing to the interaction frame defined by $\hat U_{\text{ctrl}}(\tau) = \mathcal T \exp(-i\int_{j-1 t_G}^{\tau} dt' \hat H_{\text{ctrl}}(t') )$, we obtain
\begin{equation}
\tilde g_{j,\vec \eta} = \hat g_j \mathcal T \exp\left( -i \int_{(j-1)t_G}^{j t_G} dt' \sum_\alpha \eta^{(\alpha)}(t') \hat A^{(\alpha)}(t') \right),
\end{equation}
where $(\hat A^{(\alpha)}(t)=\hat U_{\mathrm{ctrl}}^\dagger(t) \hat A^{(\alpha)} \hat U_{\mathrm{ctrl}}(t))$.

To twirl the noise, we simply insert random Pauli gates before and after each layer in complete analogy with  Eq.~\eqref{eq:twirled_layer}. The twirled channel is then given by averaging over the different Pauli gate sequences:
\begin{multline}
    \mathcal N_{j, \vec \eta} =\\ \frac{1}{4^n} \sum_{\mathcal P} \mathcal P \mathcal T \exp(-i \int_{(j-1)t_G}^{j t_G} dt' \sum_{\alpha} \eta^{(\alpha)}(t') \mathcal L^{(\alpha)}(t')) \mathcal P,
\end{multline}
where we defined the generator of the noisy dynamics $\mathcal L^{(\alpha)}(t)[\cdot] \equiv [\hat A^{(\alpha)}(t), \cdot]$. At this point, we can bring the Pauli superoperators $\mathcal P$ inside the exponential and expand the average in a cumulant expansion as was done in the supplemental material of Ref.~\cite{brillant_randomized_2025}. Here, we only keep the second cumulant, which amounts to neglecting fourth order errors occuring during a single gate time $t_G$. Importantly, higher order terms corresponding to correlations between different gates are accurately captured in this approximation, making it non-perturbative in the correlation time of the noise. Since we are interested in noise with correlation times that are large compared with the gate time $t_G$, we replace the random variables $\eta^{(\alpha)}(t)$ with their average during a given gate time $\theta_j^{(\alpha)}/t_G = \frac{1}{t_G}\int_{(j-1)t_G}^{j t_G} dt \eta^{(\alpha)}(t)$. In doing so, we obtain the twirled noise channel corresponding to the $j$th Clifford gate:
\begin{widetext}
\begin{align}
    \mathcal N_{j}(\theta_j) &= \exp(-\frac{1}{2} \sum_{\alpha,\alpha'} \theta_j^{(\alpha)} \theta_j^{(\alpha')} \mathcal K^{(\alpha,\alpha')}_{\text{avg,j}}), \label{noise_superop}\\
    \mathcal K_{\text{avg,j}}^{(\alpha, \alpha')} &= \frac{1}{4^n} \sum_{\mathcal P} \mathcal P \frac{1}{t_G^2}\int_{(j-1)t_G}^{j t_G} dt_1 \int_{(j-1)t_G}^{t_1} dt_2 \mathcal L^{(\alpha)}(t_1) \mathcal L^{(\alpha')}(t_2) \mathcal P,
\end{align}
\end{widetext}
where we defined the generator $\mathcal K_{\text{avg,j}}^{(\alpha, \alpha')}$ of the noise channel, which explicitly depends on the implementation of the gate which is applied at the $j$th time step. Since $\mathcal K_{\text{avg,j}}^{(\alpha, \alpha')}$ is a twirled superoperator, it is diagonal in the Pauli basis. We can therefore decompose it as:
\begin{equation}
    \mathcal K_{\text{avg,j}}^{(\alpha, \alpha')} = \sum_{\mathcal P} k_{j,\mathcal P}^{(\alpha, \alpha')} \mathcal P,
\end{equation}
for some coefficients $k_{j,\mathcal P}^{(\alpha, \alpha')}$. This allows us to write the total twirled noise channel in a form that is analogous to Eq.~\eqref{approx_eigvals}. To do so, we first bring all the noise superoperators at the end of the circuit by defining the circuit-evolved Pauli superoperators:
\begin{align}
    \mathcal{P}_j = \mathcal{G}_{l} \dots \mathcal{G}_{j+1}\, 
    \mathcal{P}\, 
    \mathcal{G}_{j+1}^{-1} \dots \mathcal{G}_{l}^{-1},
\end{align}
and its eigenvalues $s_{j,\mathcal P,\hat Q} = \pm 1$ with respect to its action on the Pauli operators $\hat Q$:  $\mathcal P_j\hat Q = s_{j,\mathcal P,\hat Q} \hat Q$. This allows us to write the action of the total noise channel (acting at the end of the circuit) on a given Pauli operator as:
\begin{align}
    \mathcal N_{\text{tot}}(\vec \theta)\hat Q &= \exp(-\frac{1}{2} \sum_{\alpha,\alpha',j} \theta_j^{(\alpha)} \theta_j^{(\alpha')}  \sum_\mathcal P k^{(\alpha,\alpha')}_{j,\mathcal P} s_{j,\mathcal P,\hat Q}) \hat Q\\
    &\equiv \exp(-\frac{1}{2}  \vec \theta^T \boldsymbol{M}^{\text{(f-t)}}_{\hat Q} \vec \theta  )  \hat Q\label{total_general_channel}
\end{align}
where we defined the matrix elements: $m^{(\alpha,\alpha')}_{j,\hat Q}= \sum_\mathcal P k^{(\alpha,\alpha')}_{j,\mathcal P} s_{j,\mathcal P,\hat Q}$ and the matrix $\boldsymbol{M}^{\text{(f-t)}}_{\hat Q}$ defined by the blocks $\boldsymbol({M}^{\text{(f-t)}}_{\hat Q})^{(\alpha, \alpha')} = \mathrm{diag}((m_{1,\hat Q}^{\text{(f-t)}})^{(\alpha,\alpha')}, \dots, (m_{l,\hat Q}^{\text{(f-t)}})^{(\alpha,\alpha')})$. We note that the form of Eq.~\eqref{total_general_channel} is the same as for the simpler, Z-axis only, dephasing noise model seen in the previous section, with the difference that the matrix encoding the effect of the noise on the circuit $\boldsymbol{M}_{\hat Q}^{\text{(f-t)}}$ now also depends on the way that the gates are implemented and is therefore significantly more complicated. Nevertheless, we can still derive bounds analogous to \eqref{AGF_bound}.

\subsection{Circuit fidelity bounds for finite-duration gates}
We now turn to the derivation of upper and lower bounds on the process fidelity
in the presence of finite-duration gates. 

The process fidelity can be written as a sum of terms of the form
\begin{equation}
    \lambda
    = \frac{1}{\sqrt{\det(\boldsymbol{1} + \mathbf{M}\boldsymbol{\Sigma})}},
    \label{eigvals_ft}
\end{equation}
as shown in Eq.~\eqref{eigvals}. 
Here $\boldsymbol{\Sigma}$ denotes the covariance matrix of the noise variables, 
and $\mathbf{M}$ is one of the matrices $\boldsymbol{M}_{\hat Q}^{\text{(f-t)}}$, 
which is positive semidefinite. 
Bounding the fidelity therefore reduces to bounding the determinant appearing
in Eq.~\eqref{eigvals_ft}.

\subsection{Lower bound}

The matrix $\mathbf{M}$ has a block structure indexed by pairs of qubits,
\begin{equation}
\mathbf{M} =
\begin{pmatrix}
\mathbf{M}^{(1,1)} & \cdots & \mathbf{M}^{(1,n)} \\
\vdots & \ddots & \vdots \\
\mathbf{M}^{(n,1)} & \cdots & \mathbf{M}^{(n,n)}
\end{pmatrix},
\end{equation}
where each block is diagonal in the layer index.
The covariance matrix can be written in an analogous form,
\begin{equation}
\mathbf{\Sigma} =
\begin{pmatrix}
\mathbf{\Sigma}^{(1,1)} & \cdots & \mathbf{\Sigma}^{(1,n)} \\
\vdots & \ddots & \vdots \\
\mathbf{\Sigma}^{(n,1)} & \cdots & \mathbf{\Sigma}^{(n,n)}
\end{pmatrix}.
\end{equation}

It is convenient to reorder the indices such that the matrices take a block structure with blocks labeled by the gate layer. This can be achieved by introducing a permutation
matrix $\mathbf{P}$ such that
\begin{equation}
\tilde{\mathbf{M}} = \mathbf{P}\mathbf{M}\mathbf{P}^T .
\end{equation}
In this representation $\tilde{\mathbf{M}}$ becomes block diagonal,
\begin{equation}
\tilde{\mathbf{M}} = \bigoplus_j \tilde{\mathbf{M}}_j ,
\end{equation}
where each block corresponds to a fixed gate layer,
\begin{equation}
\tilde{\mathbf{M}}_j =
\begin{pmatrix}
(\mathbf{M}^{(1,1)})_{jj} & \cdots & (\mathbf{M}^{(1,n)})_{jj} \\
\vdots & \ddots & \vdots \\
(\mathbf{M}^{(n,1)})_{jj} & \cdots & (\mathbf{M}^{(n,n)})_{jj}
\end{pmatrix}.
\end{equation}

Applying the same permutation to the covariance matrix gives
\begin{equation}
\tilde{\mathbf{\Sigma}} =
\mathbf{P}\mathbf{\Sigma}\mathbf{P}^T
=
\begin{pmatrix}
\tilde{\mathbf{\Sigma}}_{1,1} & \cdots & \tilde{\mathbf{\Sigma}}_{1,n} \\
\vdots & \ddots & \vdots \\
\tilde{\mathbf{\Sigma}}_{n,1} & \cdots & \tilde{\mathbf{\Sigma}}_{n,n}
\end{pmatrix},
\end{equation}
with blocks
\begin{equation}
\tilde{\mathbf{\Sigma}}_{j,j'} =
\begin{pmatrix}
(\mathbf{\Sigma}^{(1,1)})_{j,j'} & \cdots & (\mathbf{\Sigma}^{(1,n)})_{j,j'} \\
\vdots & \ddots & \vdots \\
(\mathbf{\Sigma}^{(n,1)})_{j,j'} & \cdots & (\mathbf{\Sigma}^{(n,n)})_{j,j'}
\end{pmatrix}.
\end{equation}

Since permutation matrices satisfy $\mathbf{P}\mathbf{P}^T=\mathbf{I}$,
the determinant appearing in Eq.~\eqref{eigvals_ft} is invariant under this transformation:
\begin{equation}
\det(\mathbf{1}+\mathbf{M}\mathbf{\Sigma})
=
\det(\mathbf{1}+\tilde{\mathbf{M}}\tilde{\mathbf{\Sigma}}).
\end{equation}

Because $\mathbf{M}$ is positive semidefinite,
$\tilde{\mathbf{M}}$ is also positive semidefinite and therefore admits a square root.
We may thus write
\begin{equation}
\det(\mathbf{1}+\tilde{\mathbf{M}}\tilde{\mathbf{\Sigma}})
=
\det\!\left(
\mathbf{1}
+
\tilde{\mathbf{M}}^{1/2}
\tilde{\mathbf{\Sigma}}
\tilde{\mathbf{M}}^{1/2}
\right).
\end{equation}

The matrix $\tilde{\mathbf{M}}^{1/2}$ inherits the block diagonal structure
\begin{equation}
\tilde{\mathbf{M}}^{1/2} = \bigoplus_j \tilde{\mathbf{M}}_j^{1/2}.
\end{equation}

We can therefore apply Fischer's inequality, which states that
for any positive semidefinite block matrix,
the determinant is bounded by the product of the determinants of the diagonal blocks.
This yields
\begin{equation}
\det(\mathbf{1}+\tilde{\mathbf{M}}\tilde{\mathbf{\Sigma}})
\le
\prod_j
\det\!\left(
\mathbf{1}
+
\tilde{\mathbf{M}}_j^{1/2}
\tilde{\mathbf{\Sigma}}_{jj}
\tilde{\mathbf{M}}_j^{1/2}
\right).
\end{equation}

The diagonal blocks $\tilde{\mathbf{\Sigma}}_{jj}$ encode the covariance
between qubits during a single gate layer,
while all temporal correlations are contained in the off-diagonal blocks
$\tilde{\mathbf{\Sigma}}_{j j'}$ with $j\neq j'$.

If these off-diagonal blocks vanish,
the covariance matrix becomes block diagonal and the above inequality
is saturated. This corresponds to the case where the noise has
no temporal correlations.
Thus, for fixed single-layer covariances,
removing temporal correlations maximizes the determinant
and therefore minimizes the fidelity.
This establishes the desired lower bound.

\subsection{Upper bound}

We now derive an upper bound on $\lambda$.
This is equivalent to finding a lower bound on the determinant
\begin{equation}
\det(\mathbf{1}+\mathbf{M}\mathbf{\Sigma}).
\end{equation}

As before we write
\begin{equation}
\det(\mathbf{1}+\mathbf{M}\mathbf{\Sigma})
=
\det\!\left(
\mathbf{1}
+
\mathbf{M}^{1/2}\mathbf{\Sigma}\mathbf{M}^{1/2}
\right),
\end{equation}
which is well defined since $\mathbf{M}$ is positive semidefinite.

For any positive semidefinite matrix $\mathbf{A}$,
the determinant satisfies the bound
\begin{equation}
\det(\mathbf{1}+\mathbf{A})
\ge
1 + \mathrm{tr}(\mathbf{A}).
\end{equation}
Applying this to $\mathbf{A}=\mathbf{M}^{1/2}\mathbf{\Sigma}\mathbf{M}^{1/2}$
gives
\begin{equation}
\det(\mathbf{1}+\mathbf{M}\mathbf{\Sigma})
\ge
1+\mathrm{tr}(\mathbf{M}\mathbf{\Sigma}).
\end{equation}

This bound is saturated when $\mathbf{M}\mathbf{\Sigma}$ has rank one.
In particular, if the covariance matrix $\mathbf{\Sigma}$ is rank one,
then $\mathbf{M}\mathbf{\Sigma}$ is at most rank one.
In this case the determinant reduces to
\begin{equation}
\det(\mathbf{1}+\mathbf{M}\mathbf{\Sigma})
=
1+\mathrm{tr}(\mathbf{M}\mathbf{\Sigma}),
\end{equation}
which saturates the inequality.

Thus, for fixed diagonal elements of the covariance matrix,
the determinant is minimized (and the fidelity maximized)
when the noise variables are maximally correlated,
corresponding to a rank-one covariance matrix.

We stress that to derive this upper bound, we started from a different setting than for the lower bound. Here, instead of fixing all the covariances during a single gate layer, we only fix the variances corresponding to a single layer and a single qubit. This is equivalent to what we did in the instantaneous gate situation.

\section{Twirling suppresses the quantum part of noise correlations}
\label{app:quantum-noise}

In this section, we show that twirling suppresses quantum correlations in the noise. We first derive the second order time-convolutionless master equation that has been used in Eq.~\eqref{TCL}. Our starting point is very similar as Eq.~\eqref{noisy_hamiltonian} from the previous section but replacing the noisy random variables $\eta^{(\alpha)}(t)$ by time-dependent bath operators $\hat B^{(\alpha)}(t)$ that evolve according to a bath only Hamiltonian $\hat H_B$. Importantly, we consider that the bath is Gaussian, stationary and initially in the state $\hat \rho_B$, meaning that all its statistics are determined by its two-point correlators $S^{(\alpha,\alpha')}(t_1-t_2) = \mathrm{tr}(\hat B^{(\alpha)}(t_1) \hat B^{(\alpha')}(t_2) \hat \rho_B)$. As explained in the main text, at leading order in the system-bath coupling, the quantum nature of the bath manifests itself through the imaginary part of $S^{(\alpha,\alpha')}(t)$. 

Here, we consider the Hamiltonian $\hat H_{\text{ctrl}, \vec \beta}(t)$ that governs the evolution of the system for the Clifford circuit subject to a specific realization of random twirling gates $\vec \beta$. Using these expressions, we can write the Hamiltonian corresponding to the system-bath evolution for a given sequence of Pauli gates as:
\begin{equation}
    \hat H_{\vec \beta}(t) = \hat H_{\text{ctrl}, \vec \beta}(t) + \sum_{\alpha} \hat A^{(\alpha)} \otimes \hat B^{(\alpha)}(t),
\end{equation}
We can then use standard techniques (see e.g. Ref.~\cite{lidar_lecture_2020}) to derive a time convolutionless master equation for the system dynamics, for a specific realization of the random Pauli gates $\vec \beta$. To leading order in the system-bath coupling, the master equation is simply given by:
\begin{equation}
     \dot{\hat{\rho}}_{\vec \beta}(t) =  \int_0^{t} dt_1 \left(\mathcal K_{\vec \beta}^{(\text{cl})}(t,t_1) + \mathcal K_{\vec \beta}^{(\text{qu})}(t,t_1) \right) \hat \rho_{\vec \beta}(t),
\end{equation}
with the classical and quantum memory kernels:
\begin{equation}
    \mathcal K_{\vec \beta}^{(\text{cl})}(t_1,t_2)[\cdot] = \sum_{\alpha, \alpha'} S_R^{(\alpha, \alpha')}(t_1-t_2)  [ \hat A^{(\alpha)}_{\vec \beta}(t_1), [\hat A^{(\alpha')}_{\vec \beta}(t_2), \cdot ]], 
\end{equation}
\begin{equation}
    \mathcal K_{\vec \beta}^{(\text{qu})}(t_1,t_2)[\cdot] = \sum_{\alpha, \alpha'} S_I^{(\alpha, \alpha')}(t_1-t_2)  [ \hat A^{(\alpha)}_{\vec \beta}(t_1), \{ \hat A^{(\alpha')}_{\vec \beta}(t_2), \cdot \}], \label{quantum colonel app}
\end{equation}
where $S_{R,I}^{(\alpha, \alpha')}(t)$ are respectively the real and imaginary part of the bath correlation function.

To make progress, we need to average over the random Pauli gates. To do so, we can first formally solve the master equation, to obtain the superoperator propagator:
\begin{multline}
    \mathcal U_{\vec \beta}(t) \\
    = \mathcal T \exp(-\int_0^t dt_1\int_0^{t_1} dt_2 \mathcal K^{(\text{cl})}_{\vec \beta}(t_1,t_2) +  \mathcal K^{(\text{qu})}_{\vec \beta,}(t_1,t_2)).
\end{multline}
To average over the random Pauli gates, we can then follow the same cumulant expansion as was used in Ref.~\cite{brillant_randomized_2025}. Keeping only the second cumulant in $\vec \beta$, we obtain the gate averaged superoperator propagator:
\begin{multline}
    \langle \mathcal U_{\vec \beta}(t) \rangle_{\vec \beta} \\
    \approx \exp \langle -\int_0^t dt_1\int_0^{t_1} dt_2 \mathcal K^{(\text{cl})}_{\vec \beta}(t_1,t_2) +  \mathcal K^{(\text{qu})}_{\vec \beta,}(t_1,t_2)\rangle_{\vec \beta}.
\end{multline}
To leading order, the dynamics can therefore simply be obtained by averaging the quantum and classical kernels separately, as was done in Sec.~\ref{sec:quantum-finite-gates}. 

To show that the quantum kernel vanishes under Pauli-twirling, we start from Eq.~\eqref{quantum_kernel_expanded}, which we rewrite here:
\begin{widetext}
\begin{equation}
    [ \hat A^{(\alpha)}_{\vec \beta}(t_1), \{ \hat A^{(\alpha')}_{\vec \beta}(t_2), \hat \rho \}] = \hat A^{(\alpha)}_{\vec \beta}(t_1) \hat A^{(\alpha')}_{\vec \beta}(t_2) \hat \rho - \hat \rho \hat A^{(\alpha')}_{\vec \beta}(t_2) \hat A^{(\alpha)}_{\vec \beta}(t_1) + \hat A^{(\alpha)}_{\vec \beta}(t_1) \hat \rho \hat A^{(\alpha')}_{\vec \beta}(t_2) - \hat A^{(\alpha')}_{\vec \beta}(t_2)  \hat \rho \hat A^{(\alpha)}_{\vec \beta}(t_1). \label{quantum_kernel_expanded app}
\end{equation}
\end{widetext} 
This equation correspond to the expansion of the nested commutator and anti-commutator terms in the quantum kernel Eq.~\eqref{quantum colonel app}. As we show next, each one of these terms vanishes when averaged over the random gate sequences $\vec \beta$, implying that the quantum kernel vanishes to lowest order in the system-bath coupling.

To compute the over the Pauli gates, we note that we can average over them in any particular order. Specifically, we choose to average over the one, $\hat P_{\beta_0}$, that was applied in the initial time step first. To do so, we factor it out of the noise operators $\hat A^{(\alpha)}_{\vec \beta}(t) = \hat P_{\beta_0} \tilde A_{\vec \beta}^{(\alpha)}(t) \hat P_{\beta_0}$, such that $\tilde A_{\vec \beta}^{(\alpha)}(t)$ is the time-evolved noise operator excluding the first random Pauli gate, which is therefore independent of $\hat P_{\beta_0}$. We can therefore average over $\hat P_{\beta_0}$ in Eq.~\eqref{quantum_kernel_expanded app}, which makes the contributions from the first and second terms proportional to the identity, with coefficients corresponding to $\mathrm{tr}[\tilde A^{(\alpha)}_{\vec \beta}(t_1) \tilde A^{(\alpha')}_{\vec \beta}(t_2)]$ and $\mathrm{tr}[\tilde A^{(\alpha')}_{\vec \beta}(t_2) \tilde A^{(\alpha)}_{\vec \beta}(t_1)]$ respectively. The cyclic property of the trace then ensures that both of these terms cancel each other. Similarly, this average renders the third and fourth terms of Eq.~\eqref{quantum_kernel_expanded app} diagonal in the Pauli basis. The coefficients associated with a given Pauli $\hat Q$ are $\mathrm{tr}[\tilde A^{(\alpha)}_{\vec \beta}(t_1) \hat Q] \mathrm{tr}[\tilde A^{(\alpha')}_{\vec \beta}(t_2)\hat Q]$ for both terms making the two terms cancel each other.

This result is perturbative in the system-bath coupling, meaning that for noise with correlation times that are much larger than a single gate time $t_G$, this is not expected to be accurate. Nevertheless, this is enough to show that to leading order in the system bath coupling, quantum correlations are suppressed by Pauli-twirling.

\section{Details of the numerical simulations}
\label{app:numerical-simulations}

In Fig.~\ref{fig:variance_fig}, we see that twirling improves the circuit fidelity. In this section, we show that this is the case because the random circuits that we generate do not involve uniformly random single qubit Clifford gate at each layer. We provide numerical evidence that for such circuits, the state fidelities of the twirled and bare cases are be the same on average over the random circuits.

\begin{figure}[ht]
    \centering
    \includegraphics[width=1\linewidth]{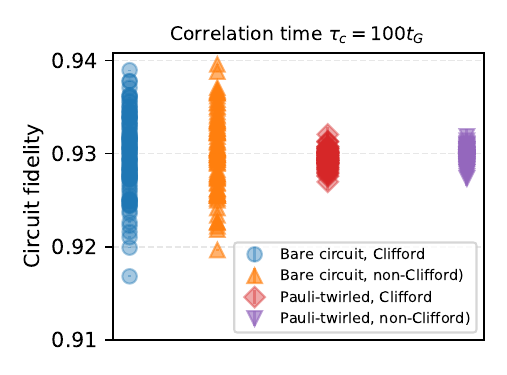}
\caption{Circuit fidelity for $100$ random Clifford and non-Clifford circuit. We consider the Pauli-twirled and bare circuit fidelities. The number of qubits is $16$ and the depth is $16$ two-qubit gate layers. The circuits are generated using the same procedure as in Fig.~\ref{fig:variance_fig} with the addition that uniformly random single qubit Clifford gates are applied between each two-qubit gate layers. Each circuit is subject to a noise with a covariance matrix Eq.~\eqref{corr_func} with correlation time is $\tau = 100t_G$ and $\sigma = 0.035$. We see that twirling reduces the variance, but does not change the average in this case.}
    \label{fig:single_qubit_cliff_sims}
\end{figure}

To do so, we simulate random brickwork circuits such as in Fig.~\ref{fig:variance_fig}, but where we insert random layers of single-qubit Clifford gates between each two-qubit gate layers. We analyze $100$ such random circuits subject to noise with covariance matrix Eq.~\eqref{corr_func} with correlation time $\tau_c = 100 t_G$ and $\sigma = 0.035$ and show the circuit fidelities in Fig.~\ref{fig:single_qubit_cliff_sims}. We observe that on average both the twirled and bare cases lead to the same circuit fidelity. We understand this because in the twirled case, the inserted Pauli gates do not add additional randomization to the circuits. This is the case because Pauli gates are a subset of Clifford gates, and therefore applying random Clifford gates already randomizes the circuits. However, we observe that Pauli-twirling does have an effect: it reduces the circuit to circuit variance in the fidelity. This is because twirling not only requires the addition of the random Pauli gates, but also to average over many random circuits. This averaging procedure reduces the circuit to circuit variance.

\clearpage

\bibliography{references}% Produces the bibliography via BibTeX.

\end{document}